\documentclass[12pt]{article}
\usepackage{xcolor}
\usepackage{graphicx}
\usepackage{pgfplots}
\usepackage[backend=biber]{biblatex}
\usepackage{listings}
\usepackage{amsmath}
\usepackage{booktabs}
\usepackage{hyperref}
\usepackage{tikz}
\usepackage{float}
\usepackage{adjustbox}
\usepackage{xspace}
\usetikzlibrary{positioning, arrows.meta}
\usepackage{xcolor} 

\usepackage{amsfonts}

\newcommand{\lambdafeedback}{\emph{Lambda Feedback}\xspace}
\newcommand{\imperial}{Imperial College London\xspace}

\makeatletter
\renewcommand\@fnsymbol[1]{\@arabic\c@footnote}
\makeatother

\title{Automated Feedback Generation for Undergraduate Mathematics: \\
Development and Evaluation of an AI Teaching Assistant}

\author{
Aron Gohr\thanks{Independent Researcher. Part of this work was completed while this author was at Imperial College London as a visiting academic. Email: aron.gohr\_at\_gmail.com}
\and 
Marie-Amelie Lawn\thanks{Imperial College London. Email: m.lawn\_at\_imperial.ac.uk}
\and 
Kevin Gao\thanks{Imperial College London. Email: kevin.gao22\_at\_imperial.ac.uk}
\and
Inigo Serjeant\thanks{Imperial College London. Email: inigo.serjeant21\_at\_imperial.ac.uk}
\and 
Stephen Heslip\thanks{Imperial College London. Email: stephen.heslip20\_at\_imperial.ac.uk}
} 

\date{\today}

\lstdefinelanguage{json}{
  basicstyle=\normalfont\ttfamily,
  numbers=left,
  numberstyle=\scriptsize,
  stepnumber=1,
  numbersep=8pt,
  showstringspaces=false,
  breaklines=true,
  frame=lines,
  string=[s]{"}{"},
  comment=[l]{:\ },
  morecomment=[l]{:"},
  commentstyle=\color{black},
  morestring=[b]',
  stringstyle=\color{black},
  identifierstyle=\color{black},
  keywordstyle=\color{black},
}

\begin{document}
\maketitle


\begin{abstract}
Intelligent tutoring systems have long enabled automated immediate feedback on student work when it is presented in a tightly structured format and when problems are very constrained, but reliably assessing free-form mathematical reasoning remains challenging.

We present a system that processes free-form natural language input, handles a wide range of edge cases, and comments competently not only on the technical correctness of submitted proofs, but also on style and presentation issues. We discuss the advantages and disadvantages of various approaches to the evaluation of such a system, and show that by the metrics we evaluate, the quality of the feedback generated is comparable to that produced by human experts when assessing early undergraduate homework. We stress-test our system with a small set of more advanced and unusual questions, and report both significant gaps and encouraging successes in that more challenging setting.

Our system uses large language models in a modular workflow. The workflow configuration is human-readable and editable without programming knowledge, and allows some intermediate steps to be precomputed or injected by the instructor.

A version of our tool is deployed on the \imperial{} mathematics homework platform \lambdafeedback. We report also on the integration of our tool into this platform.
\end{abstract}

\tableofcontents

\section{Introduction}\label{sec:intro}

\subsection{Motivation}\label{sec:motivation}
\subsubsection{Need for scalable feedback solutions}\label{sec:scalable_feedback}
Exercises are at the heart of university-level mathematics education. However, for students transitioning from high-school mathematics, it is often unclear whether they have solved a university-level assignment successfully. Unlike schoolwork, university problems require the construction of rigorous mathematical proofs—something most students have little or no prior experience with. They must tackle problems that are far more abstract and unfamiliar, often with only a vague sense of what is expected. Even when students arrive at correct solutions, many lack the experience to write them with the clarity and precision that university standards demand. Feedback at this stage is essential: without it, many students will struggle silently or fail to develop the ability to reliably check their own work.

Despite its importance, providing timely, detailed feedback on proofs remains challenging, particularly as class sizes grow and resources become increasingly limited. Traditionally, feedback is given manually by teaching assistants (TAs), typically more advanced students familiar with the course material. This approach is labour-intensive, with turnaround times often stretching to one or two weeks—assuming resources for regular homework evaluation are available at all. Even then, feedback quality can be inconsistent, and the traditional model scales poorly. Instructors and TAs must prioritise grading and administrative tasks, leaving minimal time for individualised feedback beyond general comments or standardised solutions. Moreover, those most in need of feedback—students struggling with material or time management—may be the least likely to seek it out, risking the amplification of existing equity gaps.

Existing feedback mechanisms, such as answer keys or model solutions, offer limited support. Without immediate, adaptive feedback, students are often left guessing at the causes of their mistakes, allowing misunderstandings to persist and accumulate between assessments. Even when answer keys explain correct solutions, they rarely address the specific reasoning errors individual students make, leaving significant gaps in understanding and potentially undermining future learning and motivation.

Finally, mathematics feedback requires a particularly precise combination of technical and pedagogical elements. It must address not only the correctness of the answers, but also the logical structure, notation, and step-by-step reasoning that underlie the students' solutions. Constructive feedback involves identifying common errors, suggesting alternative strategies, and clarifying misconceptions --- each of which demands significant time and specialised expertise. Although mathematics departments have access to highly qualified teaching staff capable of providing such feedback, the volume of students makes it impractical to deliver personalised, detailed feedback at scale.

Automatic feedback generation systems that produce high-quality, helpful feedback with short turnaround times and low cost could offer a solution to these challenges.  The goal of this work is to provide evidence that such systems can be built using current AI capabilities and to discuss evaluation methodology.

\subsubsection{Importance of mathematical rigour}\label{sec:rigour}
Rigour turns intuition into reliable knowledge. By requiring that every statement follow transparently from clear definitions, students learn to trace errors, think clearly, generalise ideas, and recognise links across topics. It also concerns presentation: proofs must be not only correct, but readable—and, where possible, elegant. Because this mindset differs sharply from the procedure-oriented habits fostered in school, early undergraduates need explicit guidance. Any effective feedback system must therefore flag logical gaps, imprecise language, and needless complexity, steering students toward the standards of university mathematics.

\subsubsection{Potential and limitations of LLM-based systems in mathematics education}
\label{sec:llm_potential}

Until recently, building a system that could provide such mathematical feedback would have seemed impossible. The combination of natural language understanding, mathematical reasoning, and pedagogical insight required would have been beyond the capabilities of any automated system. However, recent advances in artificial intelligence, particularly in the form of Large Language Models (LLMs), have opened up new possibilities for addressing these challenges.

LLMs are AI systems trained on vast amounts of text data that can process and generate human-like text.\footnote{Recent leading models are not in fact only trained on text, but frequently multimodal and can process images as well as text; they are also not trained exclusively on human-generated text, but to some extent on automatically generated reasoning chains or by forms of reinforcement learning. We will still call these models LLMs for brevity.} These models have demonstrated significant capabilities in tasks requiring broad knowledge and natural language understanding, including explaining complex concepts, identifying errors in writing, and providing detailed feedback. Their processing speed (when an answer is known or can be deduced relatively easily using standard techniques from known facts) and the breadth of their knowledge are better than that of humans. The most advanced models score very high on quite difficult tests of mathematical reasoning; for instance, among models that are publicly available at the time of writing, the \texttt{o3} model by OpenAI is reported to pass more than 90 percent of the AIME exam questions\footnote{AIME is an invitational exam that is part of the qualifying process for the USA IMO team and consists of 15 difficult mathematics questions. While AIME questions typically focus on pre-university mathematics topics, they require exceptional problem-solving skills and mathematical creativity, often matching or exceeding the difficulty of typical undergraduate homework problems in terms of reasoning complexity.} and performs at the level of experts in various domains when answering questions from their own domain on the Google-Proof Question Answering (GPQA) benchmark \cite{openai2024reasoning}. GPQA is designed to be a test with objective, unique answers that require very specialised knowledge where the answers are not easily deducible from web sources.

Given the challenges in providing scalable, personalized feedback and maintaining mathematical rigour in large classes, it is natural to ask whether LLM-based systems can provide a solution. \emph{If} LLMs \emph{were} sufficiently capable, they could potentially make expert-level tutoring available for everyone all the time; the question is therefore to what extent they \emph{are} capable enough to teach university-level mathematics. 
In practice, this question resolves differently across several dimensions. Table~\ref{tab:summary_llm_potential_and_limitations_v2} summarises the core challenges in large undergraduate mathematics courses, the capabilities recent LLMs bring, and the design responses we test in this project.

\begin{table}[H]
\centering
\scriptsize
\renewcommand{\arraystretch}{2}
\begin{adjustbox}{width=\textwidth}
\begin{tabular}{|p{3.5cm}|p{3.5cm}|p{3.5cm}|}
\hline
\textbf{Problem and Need} & \textbf{LLM Potential} & \textbf{Our Approach} \\
\hline
Lack of timely, personalised feedback. \newline
\textit{Need: Immediate, adaptive responses.} &
\textbf{+} Fast, personalised answers \newline
\textbf{–} May hallucinate. &
Careful prompt engineering, structure workflow to reduce errors.\\
\hline
Students struggle with proof-writing and rigour. \newline
\textit{Need: Help with logical structure and clarity.} &
\textbf{+} Good at getting the intent of unclear writing, some logical reasoning capability, broad knowledge base \newline
\textbf{–} Often vague or too generic, sycophancy. &
Decompose into subtasks; prompt for clarity, structure, and rigour-supporting feedback. Avoid both overly flattering and hypercritical feedback. \\
\hline
Large class sizes, limited teaching staff. \newline
\textit{Need: Scalable and cost-efficient support.} &
\textbf{+} Scalable at low marginal cost compared to humans \newline
\textbf{–} Budgets still constrain deployment of top-tier models. &
Use strong models selectively; optimise cost-quality balance for large deployments. \\
\hline
Equity and accessibility concerns. \newline
\textit{Need: Multilingual, always-available guidance.} &
\textbf{+} 24/7 access, multilingual output \newline
\textbf{–} Performance can vary with language, may perpetuate biases in training data &
Evaluate multilingual output; tune for clarity, equity, inclusivity. Mitigate unsafe outputs.\\
\hline\hline
\multicolumn{1}{|l|}{\textit{System design consideration}} &
Models are frequently updated; how to keep up with that? &
Structure the system modularly to automatically benefit from model improvements without major redesign. \\
\hline
\end{tabular}
\end{adjustbox}
\caption{Summary of educational challenges, LLM capabilities, and system design strategies}\label{tab:summary_llm_potential_and_limitations_v2}
\end{table}

\subsection{Related Work}\label{sec:related}

We now review prior work on automated feedback generation in mathematics and the natural sciences. While not exhaustive, this overview highlights systems and approaches conceptually related to our setting.

\subsubsection{Early Intelligent Tutoring Systems}\label{sec:related_its}

\paragraph{Expert-system-based Approaches}
Early systems used expert systems —collections of rules that model a tutor's reasoning in a given domain—to deliver timely, personalized feedback. Traditional Intelligent Tutoring Systems (ITS) of this kind were developed over decades and applied in various fields, including mathematics. Examples include Andes~\cite{gertner2000andes,vanlehn2005andes} and Cognitive Tutor~\cite{koedinger1997intelligent}, which guided students through step-by-step problem solving in physics and algebra, respectively.

Among these, Andes is most relevant to our work. Designed for undergraduate physics courses at the US Naval Academy, it combined a rule-based expert system, a graphical interface for solution entry, and a computer algebra system for verifying steps. It also checked units and displayed immediate error messages. On demand, it could diagnose mistakes or suggest next steps.

An early version (Andes 1) included a Bayesian student model to estimate conceptual understanding, but this was later removed because it provided little actionable information.

Andes was extensively evaluated and shown to improve learning outcomes~\cite{vanlehn2005andes}. However, expert-system-based ITSs proved difficult to scale due to several limitations:

\begin{itemize}
    \item High development and maintenance costs.
    \item Adding new problems required both domain experts and knowledge engineers.
    \item Each change demanded regression testing to ensure newly introduced rules did not have unintended side-effects ~\cite{vanlehn2005andes}.
    \item Dependence on structured graphical interfaces to constrain user input.
    \item Like other expert systems, these systems tended to be brittle. Edge cases had generally to be foreseen by the designer to be handled correctly.
    \item They were also unable to deal with free-form input.
\end{itemize}

\subsubsection{Mathematical Proof Assistants}
Another approach to teaching mathematical proof writing involves formal verification tools, which check completed proofs. We use the term \emph{proof assistant} for those that also support proof construction.

We first outline their capabilities and limitations, and then survey concrete educational applications.

\paragraph{Formal Verification}

Proof assistants were originally developed for the \emph{formal verification} of mathematical results, not education. Verification translates theorems and proofs into a formal language where each step can be reliably checked by a computer program. This process requires detail far beyond standard exposition—akin to implementing an algorithm in a low-level language—but offers practically absolute certainty of correctness, provided that the claim to be proven has been correctly formalised.

Despite the high effort required, several major results have been formalised, including the Feit–Thompson theorem~\cite{gonthier2013machine} and the Kepler conjecture~\cite{hales2017formal}. Formal methods are also used in software verification, for example in the verification of the seL4 microkernel~\cite{klein2009sel4}.

\paragraph{Limitations for Feedback Generation}
Despite their strengths, proof assistants remain rare in mathematics instruction. 

Major limitations include:
\begin{itemize}
  
    \item They require formal syntax and offer little support for natural mathematical writing.
    \item The level of detail exceeds that used even by professional mathematicians.
    \item Most instructors lack the expertise to use them.
    \item They check correctness but not clarity, simplicity, or elegance.
    \item For some exercises in undergraduate mathematics, obtaining a version of the question that \emph{can} even be formalised \emph{already constitutes a large part of the conceptual work}. In our test set, this is the case for instance for some of the geometry questions. These are natural-language questions that present a mathematical problem in the form of a text task, and the student has to use their understanding of the situation at hand to construct a purely mathematical question that can then be solved. Proof assistants cannot check the first step, which can be highly nontrivial. 
\end{itemize}

These limitations affect pedagogical value:
\begin{itemize}
  
    \item Learning the formal verification language creates an additional barrier of entry.
    \item Proofs may be formally correct but fail to convey insight, while insightful but flawed ones receive no recognition.
    \item Overly complex but valid proofs receive no critique on efficiency or clarity.
    \item Partial solutions with promising ideas receive little or no positive feedback.
    \item Built-in tactics may solve subgoals students cannot do by hand.
\end{itemize}

Some mitigations exist: for example, instructors can restrict tactic use or limit background results. However, generally it is fair to say that the commonly used mitigations do not address all concerns, in particular the lack of guidance on clear expression of mathematical ideas.

\paragraph{Existing Educational Applications}
Despite these issues, proof assistants remain a promising tool for feedback generation. One approach is to create constrained, gamified environments with a limited set of tactics and background results tailored to the course. A well-known example is the \emph{Natural Number Game} by Buzzard~\cite{natural_number_game,lean4game}, which teaches elementary number theory in Lean. Several similar games now exist.

Massot~\cite{massot2024teaching} proposes a version of Lean tailored for undergraduate education, featuring: (1) a syntax layer close to natural mathematical writing, and (2) a curated tactic library suitable for undergraduate-style proofs. 

Bartzia et al.\ compare six proof assistants (Lean, Isabelle, Coq, Lurch, Edukera, D$\forall\exists$duction) by solving the same elementary set theory exercise in each, analysing the process and reflecting on different purposes of proof.

Iannone and Thoma~\cite{iannone2024interactive} explored Lean (v3) in a first-year proof course. Students were introduced via a voluntary workshop; 99 completed surveys and 37 participated in interviews. Student responses were mixed. Some found that Lean enhanced their understanding of rigour. Others cited syntax difficulties, unnatural proof style, and time constraints as barriers to engagement.

A follow-up study by the same authors~\cite{thoma2021learning} analysed student proof writing in more detail. They found improved understanding overall, though it was unclear whether this was due to Lean itself or the increased time spent practising.

\subsubsection{Large Language Models}

\paragraph{Preliminaries}
Brown et al.~\cite{brown2020languagemodelsfewshotlearners} showed that LLMs trained on text prediction do more than replicate statistical patterns—they exhibit surprising capabilities in zero- and few-shot learning. This means they can solve novel tasks from minimal examples or natural language instructions.

GPT-3, introduced in that work, produced human-like responses across many domains, including basic programming and simple mathematics. Given that such models are trained on diverse educational content and can follow complex instructions, it is natural to ask whether they could support teaching. However, one problem with LLMs is that they have a tendency to make up a plausible-sounding answer when they cannot reliably answer. This made early LLMs largely unsuitable for educational applications.

\paragraph{Mathematical Reasoning Development}
Early models such as GPT-3~\cite{brown2020languagemodelsfewshotlearners} performed poorly on mathematical reasoning tasks. This improved significantly with GPT-4~\cite{openai2023gpt4systemcard}, which was the first general-purpose LLM to approach 50\% accuracy on the MATH benchmark~\cite{DBLP:journals/corr/abs-2103-03874}—achieving just over 50\% with Python support~\cite{pmlr-v202-gao23f} and around 42\% without~\cite{bubeck2023sparks}.\footnote{The Minerva model, trained specifically for mathematics, reached 50\% on MATH slightly earlier~\cite{paperswithcode_math_word_problem_solving}.} Since then, LLM performance has steadily improved~\cite{paperswithcode_math_word_problem_solving}, as has the range of benchmarks. While early tests such as MATH focused on difficult high-school problems, these are now largely saturated~\cite{plaat2024reasoning,sessler2024benchmarking}. The strongest models also perform well on more advanced problems, including those from the American Invitational Mathematics Examination (AIME)~\cite{OpenAI_LearningToReason}.

\paragraph{Empirical Trials}
LLMs have been trialled in various educational contexts, mostly at the high school level. Two noteworthy examples are LearnLM-Tutor by Jurenka et al.~\cite{jurenka2024towards} and the CoMTA dataset by Miller and Dicerbo~\cite{miller2024llm}.

LearnLM-Tutor was built by fine-tuning Google's Gemini 1.0 model~\cite{geminiteam2024geminifamilyhighlycapable}, with input from educators and a focus on pedagogical quality and safety. It was tested in Arizona State University's Study Hall and is available via Google AIStudio, based on Gemini 1.5 Pro.

Miller and Dicerbo's CoMTA dataset comprises 188 dialogues across educational levels and evaluates LLMs' ability to detect and respond to student mistakes. Using GPT-4-turbo for evaluation, they found GPT-4o-2024-05-13 achieved the highest accuracy (78.3\%), though performance varied by topic. Prompting was minimal: LLMs were simply instructed to act as helpful tutors.

\paragraph{Datasets for Education}
Few datasets are tailored to evaluating automated educational feedback. A notable example is the work of Frieder et al.~\cite{frieder2023mathematical}, who collected GPT-3.5-generated solutions to a wide range of math problems, annotated by expert graders. This enables benchmarking of autograder performance across both high school and undergraduate levels.

\paragraph{Higher Education Applications in Programming}
Automated software development tools such as OpenAI's Codex, though designed to assist programmers, can also be used in educational settings. A small but growing literature explores their use for teaching, primarily in computer science for practical programming exercises; see for instance \cite{leinonen2023comparing,kazemitabaar2023studying,denny2024computing,sarsa2022automatic}.

\subsubsection{Our Approach}

Our work builds on these developments by designing an AI teaching assistant specifically for undergraduate mathematics. Like the previously mentioned intelligent tutoring system Andes, our system is intended to be minimally invasive—requiring minimal changes to students' or teachers' existing workflows. We aim to support a broad range of mathematical reasoning tasks, similar to proof assistants, but without requiring formalised input.

Unlike prior systems, we focus not only on helping students construct correct solutions, but also on improving the clarity and structure of their exposition. In this way, we aim to overcome key limitations of earlier approaches, which were often difficult to maintain, tied to rigid interfaces, restricted to narrow domains, or burdened with steep learning curves.

\subsection{Main Contributions}\label{sec:main_contributions}
\paragraph{Overview}
This paper makes five core contributions: (1) we discuss design criteria for an ITS aimed at undergraduate mathematics; (2) we introduce an evaluation methodology to assess whether a system meets these criteria; (3) we design a modular, reconfigurable ITS based on large language models; (4) we empirically evaluate our tutoring system both on a realistic early undergraduate problem and answer dataset we constructed and on a set of stress testing questions designed to test the limits of the system's capabilities; (5) we report lessons learned from integrating the tutoring system into a live learning platform. We briefly summarise each below.

\paragraph{Design Criteria for an Undergraduate Mathematics ITS}
We propose a set of minimal design criteria for intelligent tutoring systems aimed at undergraduate mathematics. These are derived from practical design goals.

We define design criteria as properties testable at development time. They are minimal in that some goals cannot be evaluated pre-deployment and do not include advanced capabilities like multimodal input or long-turn dialogue. Our formulation allows flexibility in how each criterion is assessed.

\paragraph{Design of a Modular Feedback Generation System}
We implement a modular feedback system based on LLMs, designed to meet the above criteria for undergraduate-level mathematics. To ensure accuracy, feedback generation is decomposed into discrete tasks that the LLM can execute reliably. The system is reconfigurable through editable JSON files containing English-language instructions.\footnote{We expect this would generalise to other languages, but have not tested this.} Some criteria (e.g. safety, scalability) are satisfied directly by design; others depend on configuration and are assessed through evaluation.

\paragraph{Evaluation of the Feedback Generation System}
We evaluate the system across multiple configurations (4 LLMs and 5 workflows) using two new datasets developed for this work. Our primary evaluation approach translates system feedback into numerical grades and correlates these with human expert grades. To ensure robustness, we employ a mildly adversarial grading methodology: multiple lightweight LLMs grade each piece of feedback, and we report the worst-case performance. We also evaluate feedback conciseness through word counting and conduct qualitative analysis through manual inspection. Some design goals (safety, reconfigurability) we try to meet by design rather than empirical testing. We discuss quantitative and qualitative results as well as methodological limitations. In our assessment, the system meets the design criteria when paired with a strong LLM and suitable configuration and when used in a context that is not mathematically too advanced.

\paragraph{Integration into a Feedback Platform}
We report on integrating our tool into a live feedback platform originally designed for fixed-format responses (e.g. multiple choice). Several challenges emerged:
\begin{enumerate}
\item \textbf{Customisation Challenges.} Because mathematical notations and conventions vary, the system needs to know what conventions apply to each question. This requires giving the model context that students would have from the course. We embedded additional context into each question as needed, but of course this means that the question as it is seen by the model differs from what is displayed to the student, and crucially also to the instructor unless they specifically look at the question configuration.

\item\textbf{Technical Integration.} The platform assumed fixed-format responses; our prototype used free-form inputs. In our trial deployment, we used this to our advantage by fixing the question part of our inputs and letting a reasoning LLM generate model answers, which improved the accuracy of feedback without additional online inference cost.

\item \textbf{No Image Support.} As not all students were fluent in \LaTeX, students submitted plain text. While in our own testing we found that the underlying language models are able to handle various non-standard forms of mathematical writing in plain text, not knowing \LaTeX{} does create a barrier to using the system. Supporting handwritten input would have solved this, but integration with the handwriting recognition facilities used by \lambdafeedback was possible only in a later iteration of our test deployment than the one here reported on.

\item \textbf{Cost and Access.} While the API connection was stable, budget limits that were initially too strict led to short outages, interrupting student access and affecting usability.
\end{enumerate}

\section{System Design}\label{sec:system_design}
\subsection{Design Goals}\label{sec:design_goals}

In this subsection, we explain the design goals of our system. Subsequently, we will explain how we try to meet each of these goals, and perform evaluations where possible. Our justification for the design goals comes partly from our understanding of both success and failure modes of mathematical teaching in humans, and partly from working with early versions of our tutor itself and observing the ways \emph{it} fails or succeeds.

\subsubsection{User Interaction}
\paragraph{Ease of use} Students often begin their mathematical learning without previously encountering \LaTeX, hence it would be restrictive to immediately require these students to correctly typeset solutions. As such, the system should be resilient to receiving free-form plain text inputs, despite this leading to ambiguity in notation. At the same time, an ability to process handwriting and other forms of multimodal input (e.g. drawings, speech) would be highly desirable. These goals should be achieved while keeping the user interface simple and intuitive.



\paragraph{Understanding malformed content} The system should make at least a human-equivalent effort to understand the thinking behind poorly expressed submissions. Ideally, a poorly written submission that contains correct thinking should still receive critical feedback -- and if it is unclear to the extent that further thinking would be required, may receive critical feedback that is almost as harsh as a solution that is completely wrong.

\subsubsection{Dealing With Knowledge Gaps}

\paragraph{Feedback given should be mathematically correct} Hallucination or failures of logical reasoning are a known problem of LLMs, especially of non-reasoning models.\footnote{To be more precise, reasoning models are certainly better at logic than non-reasoning models. For resistance to hallucination, depending on the type of question asked, an advantage for reasoning models is, however, not as clear-cut \cite{openai2025o3o4systemcard}.} While an argument \emph{could} be made that there is a productive type of tutor error (for instance where a tutor gives erroneous feedback, the student catches the error, and learns to think critically), most feedback that states mathematical falsehoods is in our view not desirable. We therefore aim to minimise erroneous feedback.

\paragraph{Advice given should not be hypercritical} On the other hand, not all errors are created equal. Clerical errors have a different level of significance than conceptual errors, and even when there are multiple real errors in a submission, it seems reasonable to think that sometimes a student will be better served by a tutor who does not highlight \emph{every} problem they have found in the submission, but who identifies the \emph{main} or \emph{most fundamental} conceptual problem in the submission and helps the student avoid at least that type of error in the future. In short: feedback should be as concise as possible, and prioritise appropriately.

\paragraph{Feedback should not be authoritative when the system has no reason to be sure} When the system does not understand a question, is unable to say whether a solution is correct, or is unable to solve a question itself, it would be desirable if the system said so.

\subsubsection{Maintenance Considerations} 

\paragraph{Reconfigurability} Teachers should be able to reconfigure the system, at least within limits, without needing computer programming skills. The type of reconfiguration that a teaching academic should be able to perform on their own includes use cases like the following:
\begin{enumerate}
\item Supply context from the course that tells the system what tools the student is allowed to use and what level of detail is required.
\item Adjust choice of models in workflows to improve mathematical skill level or reduce cost.
\item Adjust some instructions within the analysis workflow to adapt the system to their use cases.
\end{enumerate}

Other steps, for instance integration of the tutor into a pre-existing e-learning platform that may exist at a given institution, or integration of specialised tools into workflows, are expected to be more complicated. We expect that these steps will require deeper technical expertise than the reconfiguration options mentioned above.

\paragraph{Costs} The system should allow tracking and limiting of API costs, to the extent that models on paid APIs are used.

\paragraph{Model agility} The system should be kept flexible with regards to the models used, both to allow for reconfiguration but also to allow switching to locally run models if a deployment requires it (for instance, for privacy reasons).

\subsubsection{Safety}
\paragraph{Moderation of harmful content} Filter mechanisms should be used to recognise harmful content or prompting that tries to elicit harmful content and moderate outputs appropriately.

\paragraph{Constructive behaviour} Criticism should always remain factual and be offered constructively, ideally even if the student were to try to provoke the system into giving non-constructive feedback.

\subsection{Architecture}\label{sec:architecture}

\subsubsection{System Overview}

Our automated feedback system is built around a modular architecture designed to be reconfigurable and to encourage the design of efficient workflows. In this paper, we describe two main instances of our system:

\begin{enumerate}
\item The research prototype, which was used for the majority of the experiments reported.
\item A version of the system that we deployed experimentally with the \lambdafeedback system.
\end{enumerate}

The research prototype is a stand-alone Python application that manages all stages of the feedback generation process. It allows the user to submit a question and associated answer, both in text or image format, and to receive feedback on the submission. In the version we used in the work on this project, it was able to run either in a graphical user interface or process batch inputs from a directory of questions and answers. The database format for batch processing was modelled after the format of the GHOSTS dataset \cite{frieder2023mathematical}. The work here reported uses only the batch processing functionality, but a version of the graphical user interface is still available in the codebase published on github.\footnote{Our repository is available at \url{https://github.com/agohr/math_tutor}.}

The system is driven by configuration files that define the workflow as well as the models to use for each step. The configuration is stored in a human readable JSON file, which contains directives written in natural language (English, in our tests).

The deployed version reuses the same processing core and configuration system, but does not use the stand-alone user interface and instead connects to \lambdafeedback. In this version, we changed the workflows used in the research prototype in one significant respect: we restricted the question submission to the fixed questions set by the instructor (i.e. students could only get feedback on homework questions assigned in their course), and precomputed exemplary solutions for each question using a reasoning model (o1-mini) to both reduce the cost of the system and improve the quality of the feedback. However, to keep configuration simple, we kept this preprocessing step completely automated, i.e. we did not edit the precomputed exemplary solutions.

The system follows a pipeline approach, decomposing the complex task of mathematical feedback generation into discrete, manageable stages that can be individually optimised and reconfigured.
At a high level, the research prototype consists of:

\begin{enumerate}
    \item \textbf{Graphical User Interface}:  Manages interaction with the user via a web interface. Interaction in the research prototype was originally limited to single-turn question-answer interactions, although we later added the ability to engage in follow-up conversations; in this paper, we focus on the single-turn version only.
    \item \textbf{Input Processing Module}:  Handles input, including both text and images, and converts them into a unified text representation.
    \item \textbf{Core Analysis Pipeline}: Processes student submissions and synthesises analysis results into coherent feedback through configurable workflows.
    \item \textbf{Integration Layer}: Connects the system to learning platforms.
    \item \textbf{Configuration Management}: Enables workflow customisation without code changes.
\end{enumerate}

The architecture does not inherently achieve our key design goals outlined in Section~\ref{sec:design_goals}: to what extent it does so depends to a large extent on the configured workflows, and on the models used in each step. The \lambdafeedback{} integration additionally adds preprocessing scripts that reads a question set supplied in \LaTeX, computes the solutions, and stores them along with the question in a database that is compatible with \lambdafeedback{}. The \lambdafeedback{} integration also took safety into account by adding a content moderation step.

\subsubsection{Input Processing}

The research prototype accepts inputs in multiple formats:

\begin{itemize}
    \item \textbf{Text submissions}: Free-form text containing mathematical notation.
    \item \textbf{Image transcription}: Conversion of handwritten or typeset mathematics to processable text.
    \item \textbf{Contextual knowledge}: Course-specific information and constraints. 
\end{itemize}

For image inputs, we utilise Vision-Language Models (VLMs) to transcribe mathematical notation into \LaTeX{} or plain text representations. This transcription is then combined with any manually entered information to create a description of the student's submission. The system also performs a preliminary completeness check to identify truncated submissions. This is an important usability feature, since processing of a submission takes around a minute for most of our workflows, and of course the feedback is not useful if the submission is incomplete.

\subsubsection{Core Analysis Pipeline}

The core of the system is a configurable pipeline implemented as a series of directive-driven processing steps. Each step can:

\begin{enumerate}
    \item Take the output of previous steps as input.
    \item Apply an LLM-based transformation. 
    \item Produce a new output for subsequent steps.
\end{enumerate}

The directives system gives the user a simple way to express small workflow graphs with steps expressed in natural language. As each LLM call has some cost in terms of both waiting time and tokens, we expect that typical workflows will be small and that they will not need more complex control flow than our system provides; in particular, we have not implemented loops or conditionals. More complex control flow could be useful in future work, but in a setting like ours, where the system is expected to answer in almost real-time, it would only be useful if LLM inference was very fast or if some of the steps were done by components much less computationally expensive than an LLM, like a proof verifier or a computer algebra system. We think such settings should be explored in future work, but our system has only used pure-LLM based workflows and the LLMs used had response times that allow only a few steps to be executed if a return is required within the order of one or two minutes.

In many educational settings, we expect that systems such as ours would be used to give feedback on student work on fixed questions set by the instructor. In such settings, it is possible to implement some fairly deep workflow graphs while maintaining fast response times. This is the case when most of the nodes of the workflow can be precomputed because they are independent of the student submission.

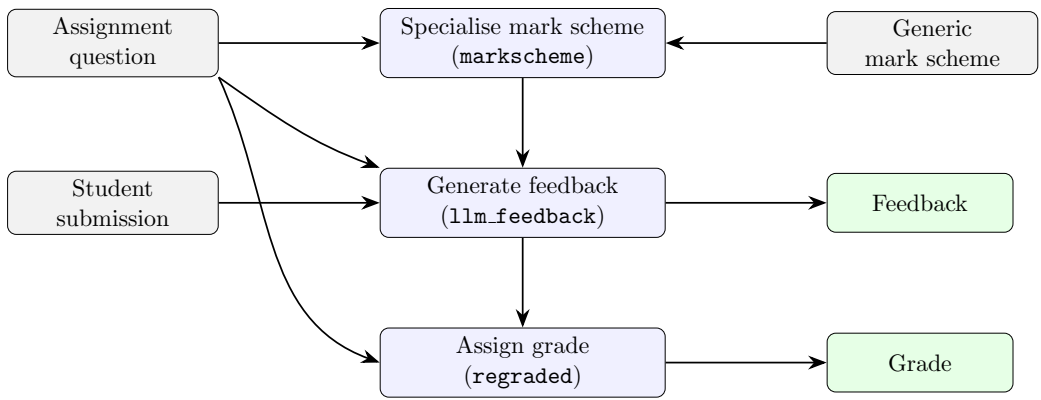
\begin{figure}[ht!]
    \centering
    \begin{adjustbox}{width=\linewidth}
    \begin{tikzpicture}[
        font=\small,
        node distance=1.45cm and 2.6cm,
        input/.style={rectangle, draw, rounded corners, fill=gray!10, minimum width=3.4cm, minimum height=0.95cm, align=center},
        process/.style={rectangle, draw, rounded corners, fill=blue!6, minimum width=4.6cm, minimum height=1.05cm, align=center},
        output/.style={rectangle, draw, rounded corners, fill=green!10, minimum width=3.0cm, minimum height=0.95cm, align=center},
        arrow/.style={-{Stealth[length=3mm]}, thick}
    ]
    \node[process] (mk) {Specialise mark scheme\\(\texttt{markscheme})};
    \node[process, below=of mk] (fb) {Generate feedback\\(\texttt{llm\_feedback})};
    \node[process, below=of fb] (grade) {Assign grade\\(\texttt{regraded})};

    \node[input, left=of mk] (question) {Assignment\\question};
    \node[input, left=of fb] (submission) {Student\\submission};
    \node[input, right=of mk] (genericms) {Generic\\mark scheme};

    \node[output, right=of fb] (fbout) {Feedback};
    \node[output, right=of grade] (gradeout) {Grade};

    \draw[arrow] (mk) -- (fb);
    \draw[arrow] (fb) -- (grade);

    \draw[arrow] (question) to[out=0, in=180] (mk);
    \draw[arrow] (genericms) to[out=180, in=0] (mk);
    \draw[arrow] (submission) to[out=0, in=180] (fb);
    \draw[arrow] (question.south east) to[out=-35, in=160] (fb.north west);
    \draw[arrow] (question.south east) to[out=-65, in=160] (grade.west);

    \draw[arrow] (fb) -- (fbout);
    \draw[arrow] (grade) -- (gradeout);
    \end{tikzpicture}
    \end{adjustbox}
    \caption{High-level view of the directive workflow configured in Listing~\ref{lst:config_example}: inputs (question, submission, generic mark scheme) feed into a small workflow graph that produces feedback, with grading as a subsequent step. In the work reported on, the \emph{grade} output is only important for evaluation purposes, and not used or generated in deployment.}
    \label{fig:pipeline_architecture}
\end{figure}

A simple example pipeline is shown in Figure~\ref{fig:pipeline_architecture}.

Listing~\ref{lst:config_example} shows how such a workflow would be configured as a directives file.
\clearpage

\begin{lstlisting}[caption={Slightly modified version of the JSON configuration for the ms\_w\_example\_final directive program (different models for different steps specified, generic mark scheme omitted for brevity; original available on github)}, label={lst:config_example}, language=json, basicstyle=\tiny\ttfamily]
{
    "context_instructions": "",
    "directives": {
        "prompt": null,
        "output": null,
        "markscheme": "You are an assistant professor of mathematics at a top university. You will be given a first-year mathematics student assignment question and a general mark scheme. Your task is to use the general mark scheme to create a specialized mark scheme that is specific to the provided assignment question. The mark scheme you output should be one paragraph and convey that rigor and correctness of mathematical argument are more important than clarity and notation when determining the quality of a solution.\n\nThe question is:\n\"\"\"\n{prompt}\n\"\"\"\n\n The general mark scheme is: \n\"\"\"\n{human_markscheme}\n\"\"\"",
        "llm_feedback": "You are an assistant professor of mathematics at a top university. You will be given an assignment question, a student's solution, and a mark scheme for the question. Your task is to evaluate the student's solution, by following the guidelines of the provided mark scheme to create helpful and concise feedback. Your feedback should be objective, do not be overly critical of good solutions with superficial errors and do not not give undue credit to poor solutions that demonstrate fundamental misunderstanding. You should only output feedback designed to support the student's learning, do not output a revised solution and withhold any grade to avoid misinterpretation. Be precise, clear and specific.\n\nThe question is:\n\"\"\"\n{prompt}\n\"\"\"\n\nThe student's solution is:\n\"\"\"\n{output}\n\"\"\"\n\nThe mark scheme is: \n\"\"\"\n{markscheme}\n\"\"\"",
        "regraded": "You are an advanced mathematics teaching assistant. Your task is to assign a grade to a student's work based solely on the question, the provided feedback, and the provided marking criteria. The grade is a number between 0 and 5, do not output anything else. \n You will now be given the question, the marking criteria and the feedback.\n\n The Question: {prompt} \n\n  The Marking Criteria written in LaTeX:\n\n While clarity and conciseness are important, correctness and mathematical rigour are the primary factors in determining a mark. Please note that marks at the extremes of the scale should be awarded where appropriate.\n\n\\paragraph{0/5 marks}: No relevant steps taken (ideas would not appear in an ideal proof); lack of understanding (e.g. alluding to unrelated results).\n\\paragraph{1/5 marks}: Some correct ideas or elements present (e.g. beginning of a proof, use of relevant definitions/theorems); mathematical argument is not valid (e.g. incorrect assumptions, misapplications of theorems)\n\\paragraph{2/5 marks}: Correct approach but would require substantial improvement (e.g. attempt at a proof); major errors present (e.g. non-trivial properties used without proof)\n\\paragraph{3/5 marks}: Correct approach with significant errors or gaps (e.g. imprecise arguments, insufficient justification)\n\\paragraph{4/5 marks}: Correct solution with minor mistakes (e.g. small logical steps omitted, arithmetic errors that do not qualitatively affect the answer)\n\\paragraph{5/5 marks}: Correct and complete solution (e.g. rigorous and well structured); minimal superfluous steps (allow some leeway, e.g. using trivial results without thorough justification).\n\n The Feedback:\n\n {llm_feedback}"
    },
    "parameters": {
        "markscheme": {"model": "gpt-5"},
        "llm_feedback": {"model": "gpt-5"},
        "regraded": {"model": "gpt-4o", "temperature": 0.0}
    }
}
\end{lstlisting}

Each directive specifies what the LLM should do, referencing outputs from previous steps in curly braces. The parameters section allows fine-grained control over which model is used for each directive and at what temperature, enabling cost-quality trade-offs.

\subsubsection{Integration Capabilities}\label{sec:integration}

The architecture supports integration with existing educational platforms by offering the core functionality of executing configurable workflows while accepting configurable input fields through a Python class interface. For our deployment with the \lambdafeedback system (see Section 5), we implemented adapters that convert between the platform's fixed-format question structure and our free-form feedback system.

Integration with a pre-existing platform required some adaptation of our expected input and output formats; we also removed image inputs for deployment, because \lambdafeedback{} had a different, already established workflow for image inputs. We also needed to add logic to handle user rate limits to avoid exceeding API budgets for the test, but this was straightforward as the platform provided a way to see how many requests had already been made by the current student user.

\section{Experimental Methods}\label{sec:methods}
\subsection{Datasets}\label{sec:datasets}
This section outlines the construction of the dataset used to evaluate our prototype, including the educational context, question selection, and preparation of student responses for testing.

\subsubsection{Context of Questions}\label{sec:context_questions}
\emph{Introduction to University Mathematics} (IUM) is a compulsory four-week module for all first-year mathematics students at \imperial. It introduces formal reasoning, logical structure, and mathematical writing—skills essential for undergraduate study. Building directly on the pre-arrival material, it serves as a unified foundation for the rest of the first-year curriculum, with an emphasis on axiomatic methods and rigorous proof techniques.

The IUM exercises are well suited to our investigation: they are self-contained, proof-based, and carefully designed to develop logical precision, while also involving substantial bridging-level computation. Their selection supports students' transition from school mathematics to the standards of university-level proof and abstraction, making them an ideal testbed for evaluating automated feedback.

\subsubsection{Content of Questions} \label{questions_content}

The IUM problem set contains over 60 exercises, rated from one to three stars in difficulty. For this study, we selected a subset of challenging and conceptually rich problems from the first two topics:

\begin{itemize}
    \item \textbf{Topic 1:} Number systems. This covers the construction of the number systems, starting with the Peano Axioms and the natural numbers, and proceeding through the formal construction of the integers and rationals, ending with the axioms of the real numbers. Students re-derive properties they have previously taken for granted, such as associativity of addition or trichotomy of order. These exercises are intentionally stripped of visual aids or heuristic shortcuts—students must reason strictly from axioms, definitions, and formal deduction. The result is a deliberate break from intuition and a rigorous introduction to abstract reasoning.

    \item \textbf{Topic 2:} Vector Geometry and Vector Calculus. This topic is more applied, but equally foundational. Students work with lines, planes, and distances in \(\mathbb{R}^3\), constructing geometric arguments in vector notation. The emphasis shifts from axiomatic reasoning to bridging geometric intuition with algebraic formulation, preparing students for later modules in linear algebra and multivariable calculus.
\end{itemize}

 \subsubsection{Generation of Solutions}
The question set was divided among three members of the team. These were experienced students who had previously completed the IUM and are co-authors of this paper. Two were in their second year of study at the time of the active part of the project, and one in their third. Each independently wrote solutions to a subset of the questions. While most of the resulting answers were of high quality and reflected strong proof attempts, they naturally varied in style, structure, and accuracy. Some solutions included inadvertent mistakes, while others were deliberately written to contain common misconceptions or conceptual errors, simulating the kinds of responses a struggling student might submit. In total, 65 solutions were generated.

\subsubsection{Constructing a Dataset for Marking and Feedback}

Our main dataset consists of four components: the questions, the student-written solutions, numerical grades from 0 to 5 assigned by three human markers (two of whom graded half the dataset and one provided an independent grading run for the whole dataset), and short written feedback aligned with undergraduate teaching standards. These were used as the benchmark for evaluating LLM-generated feedback and grading accuracy. Grades were for the final version of the dataset reconciled between the two independent human grading runs. The reconciliation process was done by one of the authors based on manual review of all disagreements.

\paragraph{Validating the Grader}

First, we needed to validate the idea that our feedback could be used to infer grades. In that way, we could test whether at least feedback \emph{identical} to the human feedback would produce high grade correlations with the human grades. This turned out to be the case.

Human feedback was used in test runs where \texttt{gpt-4o} was asked to assign a grade based solely on the feedback, with or without access to the original question. These tests assessed whether \texttt{gpt-4o} could infer a grade from realistic tutor-style feedback. As such, the feedback needed to be both informative enough to imply a mark and concise enough to remain authentic in the sense that a real tutor could afford to comment at that level of detail on the work of \emph{every} student.

Evaluation revealed that minor changes in phrasing led to significant variation in LLM-assigned grades. For example, using “correctly identifies” resulted in lower grades, even when the proof was accurate, whereas rephrasing as “proves” or “constructs” produced appropriate marks (e.g.\ a shift from 0 to 5). Accordingly, some feedback entries were reworded to test how linguistic cues influenced the model’s interpretation. However, overall correlation between LLM-assigned grades and human grades was quite high, showing that \texttt{gpt-4o} was to a large extent able to infer grades from human feedback.



\paragraph{Marking Scheme:} To ensure consistency, the marking scheme was improved over several iterations:

\begin{itemize}
  \item \textbf{Mark Scheme 1} was the initial version. It described each mark level qualitatively, using phrases such as ``some correct elements'' or ``significant mistakes''. However, it included overlapping criteria, which made the distinction between adjacent scores ambiguous.
  
  \item \textbf{Mark Scheme 2} addressed these ambiguities by introducing mutually exclusive criteria for each mark. It distinguished clearly between ``major'' and ``minor'' errors and provided both positive and negative indicators for each grade level. It also included representative examples to aid interpretation.
  
  \item \textbf{Mark Scheme 3} was a more concise formulation. It reduced each mark level to one defining strength and one limitation, improving interpretability for the LLM while preserving rigour. This version was found to improve performance in grading tasks.
  
  \item \textbf{Mark Scheme 4} further simplified the format. It retained the structure of one positive and one negative feature per grade level, resulting in short but clear instructions for each of the six marks from 0 to 5. This version was used for the final evaluation.
\end{itemize}

\subsection{Evaluation Methods}\label{sec:evaluation_methods}
Our evaluation uses different approaches to check different design goals. In the sequel, we briefly discuss our methods to evaluate how well a particular configuration of our system hits the goals discussed.
\paragraph{Design Goals Not Tested} Some of the design goals are difficult to test outside of practical deployment. We list these goals briefly below and mention to what extent we nonetheless made an effort to test for them or to make sure that they are satisfied:
\begin{enumerate}
\item Safety can ultimately only be judged in large deployments with monitoring of potentially unsafe inputs and outputs, or alternatively in adversarial testing, i.e. using jailbreaking studies. The deployment setting we target (tutoring in an undergraduate mathematics context) is in our judgment relatively low-risk to begin with. Our design decisions to allow only single-turn interactions and to filter inputs through OpenAI's standard moderation endpoints should further lower the likelihood of outputs that are directly harmful to the student or inappropriate. In our test deployment, we further reduced these problems by making the tutor available only for submission of answers to fixed homework questions set by the course instructor, and with a query limit per student.
\item Reconfigurability can likewise only be tested in deployment, ideally outside test deployments where developers can directly troubleshoot problems. The directive system does allow reprogramming the tutor in natural language, although the use of JSON to encode the directives still represents a barrier to entry.
\end{enumerate}

\paragraph{Overall Evaluation} Our main tool for overall quantitative evaluation of tutor performance is comparison of tutor feedback with the \emph{grading} of the answers submitted returned by human experts. The main rationale behind this is as follows:

\begin{enumerate}
    \item Grades implicitly contain many of the judgments that also go into feedback generation: for instance, both contain an assessment of mistakes made in a submission, of their relative severity, and also of aspects of a submission that are not strictly speaking mathematical mistakes, like unclear expression.
    \item Given a dataset of submissions, grades and short written feedback can be collected once from human experts and can then be reused when evaluating different system configurations. By contrast, direct evaluation of the feedback itself by human experts would probably provide a more reliable measure of feedback quality, but implies a large manual work factor that has to be repeated for every iteration of the system.
    \item We also wanted to be able to make use of the pre-existing dataset by Frieder et al. \cite{frieder2023mathematical}, which contains a broad set of questions and solutions that roughly resemble mathematical undergraduate homework in terms of mathematical depth, but which contains only grades and not detailed feedback on the solutions.
\end{enumerate}

Therefore, we extend all of our feedback generation workflows by a \emph{grading} step. The grading step takes the feedback and is instructed to derive a numerical grade for the submission purely from the feedback. The grading step was optimised by manually trying a small number of grading prompts and testing them on the human-generated feedback on our test submissions. Grader performance was optimised by simply picking the grading prompt that produced the highest Pearson correlation between human grades and automatically generated grades based on the human feedback.

This preliminary step of optimising the grader used the full dataset, but we do not think that this creates a data contamination risk for the experiments subsequently run because grader optimisation only used the human-generated feedback as input. This feedback is not used in the subsequent evaluation pipeline, and we settled on the final grading prompt after trying fewer than 20 possibilities, so we did not exert high optimisation pressure on the grading mechanism.

Additionally, we evaluate the conciseness of the feedback by simple word counting. A word here is just a sequence of characters separated by whitespace or punctuation marks.


\paragraph{Methodology Details for the Grading Step}
The grading step is in our study performed by an additional workflow step that is run after the feedback generation step.

The translation of feedback to a grade is an important step in our quantitative evaluation. As such, it is important to realise that it comes with some choices that must be made and which contain non-obvious tradeoffs. We briefly discuss these choices and the choices we made in our experiments here.

First, we need to decide on the rubric to use, i.e. on the general instructions that the model will follow when grading. We chose a rubric by trying a few variants and selecting among the tested variants the one that gave the most accurate translation of \emph{human-written feedback} to human grades. The rubric we chose gives short but clear instructions for each of the six grade levels from 0 to 5.

Second, we need to decide what information will be available to the model when grading. One can make a case both for just showing the grader the feedback, and for showing the grader the question and the feedback. Sometimes it is useful for the grader to know this context, because without it the importance of any problems mentioned in the feedback may be unclear; on other hand, showing the grader the question and the feedback in principle allows an extremely simple ``feedback generator'' that just copies the student submission verbatim and leaves evaluation entirely to the grader, which could be an optimal strategy if feedback generation is much harder than grading. We opted to accept this risk, but are overall agnostic on which variant would be methodically superior for the present paper. In no case did we show the grader the student submission, because this would allow the grader to ignore the feedback and assess the student's answer on its own.

Third, we need to decide \emph{which model} will be used to grade the feedback. There are two natural choices: either use the same model that was used to generate the feedback, or use a different model. We opted for the latter, because we are trying to assess the quality of the feedback generation system, and using different grading steps for each model/workflow combination could bias evaluation towards models that are capable of performing accurate grading while ignoring the feedback.

Concretely, we used a grading step that is \emph{mildly adversarial} to the feedback generation system. Essentially, after a particular model/workflow combination has generated feedback, we use a range of different grading models to grade the feedback, compute our metrics for grading accuracy with respect to reconciled human grades, and use as the true score of the model/workflow combination the \emph{worst-performing} grading model. Our selection of grading models spans a range of predominantly lightweight models: \texttt{gpt-4o}, \texttt{gpt-4.1-nano}, \texttt{gpt-5-nano}, \texttt{gpt-5-mini}, and \texttt{gpt-4o-mini}.

The rationale for doing this is that good feedback must be robust against being misunderstood by a student who struggles with the material presented. We propose that using a range of different, relatively lightweight models in the grading step captures this robustness requirement, albeit in a very rough way: most of the models used in the grading step are unlikely to be good enough at mathematics to perform the grading task without the aid of the feedback, and if they misunderstand the feedback, the likelihood is high that students might do so as well. If, on the other hand, feedback generated can be reliably translated to grades that correspond well to the judgments of human experts even by a model such as \texttt{gpt-4.1-nano}, then it is likely that students will likewise understand it correctly.

In total, we used eight different feedback generation models (\texttt{gpt-4o}, \texttt{gpt-4.1}, \texttt{gpt-4.1-mini}, \texttt{gpt-4.1-nano}, \texttt{gpt-5}, \texttt{gpt-5-mini}, \texttt{gpt-5-nano}), five different workflows, and five grading models, for a total of 200 different model/workflow/grading model combinations and 40 model/workflow combinations. Apart from the three best workflows from our workflow development, two variations of a basic workflow were included in the evaluation: both just see the question and the student submission and \emph{directly} generate feedback. The two basic workflows differ from each other in that one includes an instruction to limit feedback length to 200 words, while the other does not.

The grading prompt we used is given in Box \ref{box:grading_prompt} below.

\begin{center}
\fbox{\begin{minipage}{0.95\textwidth}
\vspace{0.2cm}
\textbf{Listing \refstepcounter{table}\thetable: Grading Prompt}\label{box:grading_prompt}
\vspace{0.2cm}

\small
\noindent You are an advanced mathematics teaching assistant. Your task is to assign a grade to a student's work based solely on the question, the provided feedback, and the provided marking criteria. The grade is a number between 0 and 5, do not output anything else. You will now be given the question, the marking criteria and the feedback.

\vspace{0.2cm}
\noindent \textbf{The Question:} \texttt{\{prompt\}}

\vspace{0.2cm}
\noindent \textbf{The Marking Criteria written in \LaTeX{}:}

\vspace{0.1cm}
\noindent While clarity and conciseness are important, correctness and mathematical rigour are the primary factors in determining a mark. Please note that marks at the extremes of the scale should be awarded where appropriate.

\paragraph{0/5 marks}: No relevant steps taken (ideas would not appear in an ideal proof); lack of understanding (e.g. alluding to unrelated results).

\paragraph{1/5 marks}: Some correct ideas or elements present (e.g. beginning of a proof, use of relevant definitions/theorems); mathematical argument is not valid (e.g. incorrect assumptions, misapplications of theorems)

\paragraph{2/5 marks}: Correct approach but would require substantial improvement (e.g. attempt at a proof); major errors present (e.g. non-trivial properties used without proof)

\paragraph{3/5 marks}: Correct approach with significant errors or gaps (e.g. imprecise arguments, insufficient justification)

\paragraph{4/5 marks}: Correct solution with minor mistakes (e.g. small logical steps omitted, arithmetic errors that do not qualitatively affect the answer)

\paragraph{5/5 marks}: Correct and complete solution (e.g. rigorous and well structured); minimal superfluous steps (allow some leeway, e.g. using trivial results without thorough justification).

\vspace{0.2cm}
\noindent \textbf{The Feedback:} \texttt{\{llm\_feedback\}}

\vspace{0.2cm}
\end{minipage}}
\end{center} 





\section{Results}\label{sec:results}
\subsection{Overview of Results}\label{sec:overview_of_results} In this section, we present the quantitative and qualitative results of our study. We first discuss quantitative results, then qualitative results, and finally, we discuss the lessons learned from our study.
The subsection on quantitative results will first discuss grade correlations between human and LLM grades, and finally the coverage and conciseness of the feedback as discussed in the previous section.

The subsection on qualitative results will discuss the findings from manual testing of the feedback generator and manual inspection of feedback generated in the automated evaluation. It will also cover a manual comparison of human and LLM feedback. Of necessity, the results of the qualitative testing will be more anecdotal than the quantitative results, as it is difficult to turn these observations into clear metrics. We do however believe that the qualitative results give important insights into the behaviour of the feedback generator, on the one hand by exercising edge cases, and on the other hand by capturing aspects of the feedback that may be important for practical teaching deployment, but hard to capture otherwise.

The subsection on lessons learned will discuss what we learned from the study in terms of design choices for the feedback generator, for the process of optimising such a system, and for evaluating these systems in the small sample size regime our study operates in.

\subsection{Quantitative Results}\label{sec:quantitative_results}
Our quantitative results are all based on measuring the correlation between grades derived from the feedback our tutoring system produces in various configurations and the human grades in our dataset. 
\paragraph{A Baseline: Agreement Between Human Graders} To establish a baseline for the LLM graders, we first compared the grades assigned by two independent human expert graders for all 65 questions and answers in our test set. We then manually reviewed all disagreements between the two human graders and reconciled them. The resulting grades were used as the human baseline against which all LLM graders were compared.

Table \ref{tab:grade_correlations_human_graders} shows the Pearson correlation, Kendall's tau, Spearman's rank correlation, percentage of exact agreement, percentage of close match (within 1 point), mean difference, mean absolute error, and mean squared error between the two human graders.

\begin{table}[H]
\centering
\caption{Correlation metrics between two human graders}
\label{tab:grade_correlations_human_graders}
\begin{tabular}{|l|c|}
\hline
\textbf{Metric} & \textbf{Value} \\
\hline
Number of questions & 65 \\
\hline
Pearson correlation & 0.4910 \\
\hline
Kendall's tau & 0.3873 \\
\hline
Spearman rank correlation & 0.4727 \\
\hline
Exact agreement (\%) & 32.3 \\
\hline
Close match (within 1 point) (\%) & 72.3 \\
\hline
Mean absolute error & 1.0769 \\
\hline
Mean squared error & 2.1231 \\
\hline
\end{tabular}
\end{table}

While there is to the best of our knowledge not much published work on inter-grader agreement for free-form undergraduate mathematics homework, the observed level of agreement is roughly in line with expectations set for instance by Moore \cite{moore2016mathematics}. Although Moore's study was too small to make quantitative comparisons to our dataset meaningful (they considered six proofs, which were graded by four mathematics professors), their qualitative observation of significant variation in grading is consistent with our findings. Moore's study also suggested as one major source of variation differing judgement on the \emph{seriousness} of a mistake; in their study, this was a greater source of variation than straightforward errors in proof assessment. The mathematical level of the proofs in Moore's study is quite close to our dataset, which makes it an interesting point of comparison.

Both of our human graders are much more closely correlated with the reconciled grades than the numbers reported in Table \ref{tab:grade_correlations_human_graders} would suggest. This is due to the fact that the reconciled grades are derived from the raw grades of the two human graders. We therefore do not report these numbers here, as they would not be meaningful.


\paragraph{Grading Accuracy of LLM Graders}\label{sec:grading_accuracy_of_llm_graders}
We will now assess the accuracy of the three best LLM graders (as judged by their Pearson correlation with the human grades on the development set) by comparing their grades to the \emph{reconciled} grades of the two human graders. Since 32 questions in our dataset were used in the prompt development of the LLM graders, they are to some degree tainted by this leakage. All prompt development was done with \texttt{gpt-4o}, so all inter-model comparisons can to some extent be assumed to be biased towards higher correlation with the \texttt{gpt-4o} grades. However, the reported grades are from independent runs of the LLM graders, so we can assume that at least processing stages after the direct processing of the question and the answer will diverge from data seen in prompt development, which should limit the biases. Also, as we already pointed out in Section \ref{sec:methods}, our main metric is the observed correlation between LLM grades and human grades after conversion of the given feedback to grades by the automated grading model out of five mostly relatively weak/fast models that performs \emph{worst} in terms of correlation with the human grades; so on balance, the reported numbers should paint a pessimistic picture of the extent to which the LLM-generated feedback correctly points out errors and weaknesses in the student's solution.

\begin{table}[ht]
\centering
\caption{Model robustness analysis: Top 5 feedback model/workflow combinations by worst-case Pearson correlation across 5 grading models.}
\label{tab:llm_robustness_best}
\resizebox{\textwidth}{!}{%
\begin{tabular}{llccccccccc}
\toprule
 & & \multicolumn{2}{c}{\textbf{Worst Pearson}} & \multicolumn{2}{c}{\textbf{Best Pearson}} & \multicolumn{2}{c}{\textbf{Worst Kendall $\tau$}} & \multicolumn{2}{c}{\textbf{Best Kendall $\tau$}} & \\
\cmidrule(lr){3-4} \cmidrule(lr){5-6} \cmidrule(lr){7-8} \cmidrule(lr){9-10}
\textbf{Workflow} & \textbf{Feedback Model} & \textbf{r} & \textbf{Grader} & \textbf{r} & \textbf{Grader} & \textbf{$\tau$} & \textbf{Grader} & \textbf{$\tau$} & \textbf{Grader} & \textbf{Avg Length} \\
\midrule
ms\_w\_example\_final & gpt-4.1-2025-04-14 & 0.572 & gpt-5-mini & 0.636 & gpt-5-nano & 0.487 & gpt-5-mini & 0.551 & gpt-5-nano & 264 \\
ms\_w\_example\_final & o4-mini & 0.521 & gpt-4.1-nano & 0.552 & gpt-4o & 0.421 & gpt-5-nano & 0.455 & gpt-5-mini & 233 \\
ms\_w\_example\_final & gpt-4.1-mini-2025-04-14 & 0.517 & gpt-4o-mini & 0.599 & gpt-4o & 0.421 & gpt-4o-mini & 0.508 & gpt-5-nano & 258 \\
baseline & gpt-4o-2024-08-06 & 0.515 & gpt-4.1-nano & 0.546 & gpt-4o & 0.420 & gpt-5-mini & 0.446 & gpt-4o & 347 \\
ms\_w\_example\_final & gpt-4o-2024-08-06 & 0.485 & gpt-4.1-nano & 0.688 & gpt-4o-mini & 0.421 & gpt-4.1-nano & 0.602 & gpt-4o-mini & 276 \\
\bottomrule
\end{tabular}%
}
\small
\textit{Note: Each feedback model/workflow combination was evaluated by 5 different grading models. Table shows the worst and best performing graders for each configuration by Pearson correlation and Kendall's tau. Avg Length measured in words.}
\end{table}

\begin{table}[ht]
\centering
\caption{Model robustness analysis: Bottom 5 feedback model/workflow combinations by worst-case Pearson correlation across 5 grading models.}
\label{tab:llm_robustness_worst}
\resizebox{\textwidth}{!}{%
\begin{tabular}{llccccccccc}
\toprule
 & & \multicolumn{2}{c}{\textbf{Worst Pearson}} & \multicolumn{2}{c}{\textbf{Best Pearson}} & \multicolumn{2}{c}{\textbf{Worst Kendall $\tau$}} & \multicolumn{2}{c}{\textbf{Best Kendall $\tau$}} & \\
\cmidrule(lr){3-4} \cmidrule(lr){5-6} \cmidrule(lr){7-8} \cmidrule(lr){9-10}
\textbf{Workflow} & \textbf{Feedback Model} & \textbf{r} & \textbf{Grader} & \textbf{r} & \textbf{Grader} & \textbf{$\tau$} & \textbf{Grader} & \textbf{$\tau$} & \textbf{Grader} & \textbf{Avg Length} \\
\midrule
use\_markscheme\_no\_example3.0 & gpt-5-mini & 0.164 & gpt-4.1-nano & 0.550 & gpt-5-nano & 0.143 & gpt-4.1-nano & 0.450 & gpt-5-nano & 333 \\
baseline\_concise & gpt-5-nano & 0.157 & gpt-4.1-nano & 0.490 & gpt-4o & 0.134 & gpt-4.1-nano & 0.417 & gpt-4o & 214 \\
use\_markscheme\_no\_example3.0 & gpt-5 & 0.112 & gpt-4.1-nano & 0.505 & gpt-4o & 0.074 & gpt-4.1-nano & 0.412 & gpt-4o & 228 \\
systematic5.0 & gpt-5 & 0.040 & gpt-4.1-nano & 0.546 & gpt-5-mini & 0.100 & gpt-4.1-nano & 0.425 & gpt-5-mini & 207 \\
systematic5.0 & gpt-5-nano & $-0.010$ & gpt-4.1-nano & 0.473 & gpt-4o & 0.020 & gpt-4.1-nano & 0.395 & gpt-4o & 229 \\
\bottomrule
\end{tabular}%
}
\small
\textit{Note: Each feedback model/workflow combination was evaluated by 5 different grading models. These configurations show high variance across graders. Avg Length measured in words.}
\end{table}

The overall picture that emerges from these comparisons is that the grading correlations between well-performing models and human graders are slightly higher than or equal to the human inter-grader agreement for the same task; furthermore, some workflow/model combinations depend heavily on a good choice of grading model to achieve reasonable correlation with the human grades. This includes cases that a priori would (based on the performance of the feedback-generating model on mathematical reasoning benchmarks)be expected to perform well, such as the \texttt{gpt-5} model with the \texttt{systematic5.0} workflow.

The dataset size prevents us from reaching stronger conclusions about the advantages and disadvantages of the different models and workflows. All in all, correlation and grade agreement as a metric are useful to recognise poor workflows and models even at this small scale, but these metrics face significant ceiling effects on this dataset due to the high inter-grader disagreement on proof assessment tasks without detailed rubrics and the relatively simple (by the standards of university-level mathematics) nature of the given problems.

\subsection{Qualitative Results}\label{sec:qualitative_results}
\subsubsection{Feedback on exemplary solutions}\label{sec:feedback_on_exemplary_solutions}


We evaluated our LLM-based feedback tool (running, for these tests, with \texttt{o4-mini} as the underlying language model and with a simple solve/feedback/grade workflow) on a set of exemplary solutions from a first-year mathematics course. We had two reasons for doing this: one, we wanted to test another use case of our tool, in this case proofreading for technical accuracy, that is similar but distinct from giving feedback to students; two, our tool is designed to provide critical feedback, and as such, it is natural to ask whether it is prone to giving negative feedback even when there is nothing to be critical about.

In the majority of cases, we found the tool's feedback was substantively correct. Most flagged issues stemmed from missing problem context: for instance, whether prime-factorisation characterisations of 
gcd/lcm were permitted (Item 1), which Peano axiom numbering was being used (Item 4), or whether prior lemmas/cancellation rules were available (Items 6, 11–12). 

We found that these ``false positives'' are largely avoidable by explicitly giving the allowed assumptions as context. We note that this context problem is difficult to address in a general chat platform, because different teaching settings differ in this regard even when the material taught is the same.

The tool was also effective at detecting genuine mismatches and errors. It correctly identified solutions proving the wrong statement (Items 2 and 7), a misleading substitution that did not yield the target claim (Item 9), and a computational mistake (Item 14). In several cases it improved exposition without altering mathematical content (Items 8, 13, 15), and it supplied missing steps when the submitted solution appealed to earlier results not shown (Items 3, 11–12). The only clearly unhelpful intervention was Item 10, where the critique hinged on treating $\pm\infty$ as numbers, contrary to the intended (non-extended) real setting. This is arguably also a context issue: in an introductory Analysis course, the ambient domain is always $\mathbb{R}$, not the extended reals, and no one would ordinarily consider 
$\pm\infty$ as numbers.

Our main takeaways are as follows. The overall performance on this early undergraduate problem set at proofreading exemplary solution was very good; several problems were found, most of them clerical. Providing minimal, machine-readable context, such as the allowed axioms, prior results, and notation/conventions, substantially reduces spurious objections. The remaining failure modes are chiefly due to convention sensitivity (for example, axiom labels) and domain specification (for instance,working in $\mathbb{R}$ versus the extended reals).

\subsubsection{Stress Testing the Tutoring System}\label{sec:stress_tests}
In this section, we present qualitative results from testing the feedback generator on various feedback generation tasks that we designed to stress-test the system. In particular, we wanted to test the system's ability to handle the following situations:

\begin{itemize}
\item Feedback on correct solutions that differ significantly from the reference solution.
\item Solutions that are strictly speaking correct, but which use overly strong mathematical machinery to solve the problem
\item Topics that are likely too advanced or specialised for the underlying language model, at least if it does not have access to the internet.
\item Multi-lingual support.
\item Questions that are undergraduate-level, but which have a solution that is slightly counter-intuitive, and which are likely not widely presented in the training data of the underlying language model.
\end{itemize}

For this qualitative testing, we designed one question of each type along with a suitable answer. We then let our tutoring system give feedback on these questions and solutions. 

In order for the format of our question database to match that of our main undergraduate mathematics dataset, each question/solution pair is in the database also accompanied by model feedback on the solution and a grade; this model feedback was automatically generated by \texttt{claude-sonnet-4-5-20250929} \cite{anthropic2025sonnet45} from the comments given below on each question. These feedback entries and the accompanying grades are not used in our analysis and they should be considered as placeholder entries. If the hard questions dataset were in future work to be extended to form a hard feedback benchmark, these entries should be replaced by high-quality human-written feedback and grades.

Finally, we manually judged the quality of the feedback. Testing was done with the \texttt{ms\_w\_example\_final} workflow and the \texttt{gpt-4.1}, \texttt{o4-mini}, and \texttt{gpt-5} models. These workflows/models were chosen because the workflow showed the most consistent results in our quantitative evaluation, and because \texttt{gpt-4.1} and \texttt{o4-mini} were in turn the most consistently well-performing \emph{models} in the same tests. The \texttt{gpt-5} model was chosen because it represents the very latest generation of reasoning models as of the time of writing. The reasoning models were again run at the \texttt{medium} reasoning effort level that was also used on the main undergraduate mathematics dataset.

The rest of the section presents the results of this testing. We devote one paragraph to each of the questions, and show the question, solution, and feedback in an itemised environment in each case. At the end of the present section, we then give a summary evaluation of the results of these tests.

Full traces of the relevant workflow executions reported here can be found in our github repository. The first paragraph gives a qualitative overview of all results.

\paragraph{Overall results} All models had complete failure cases on the stress testing set, but the extent of failure observed differed wildly from one model to another. \texttt{gpt-4.1} failed on all five questions, while \texttt{o4-mini} and \texttt{gpt-5} showed good performance on two out of five tested questions. For the multilingual question, the foreign-language aspect did not seem to be a problem, but the models all opted to answer in English. Notably, the workflow \texttt{systematic5.0} performed very well on four of the five test questions when run with \texttt{gpt-5} and pushed back against the wrong answer given in test question 3 even though it failed to point the student towards the correct solution; we do not discuss it here, but the relevant experimental outputs can be found on our github. It is plausible that the careful analysis this workflow performs of the student answer before writing feedback allows it to perform better on questions the model cannot on its own find the desired answers for.

In many of the shown failure cases, mitigations are probably possible by precomputation of workflow steps that do not depend on the student answer. Systematic study of this question would require a larger test set of hard feedback problems and we therefore leave it to future work.

\paragraph{Feedback on correct solutions that differ significantly from the reference solution} To test what happens when the feedback generation encounters a correct solution that differs a lot from what it expects, we designed a question that can easily be solved with tools that are from an undergraduate-level point of view relatively heavy, but which also has a perfectly elementary solution. Our hope was that the language model would opt to produce a reference solution for itself that uses the heavy machinery, and wanted to see how well it would do at giving feedback on the elementary solution.

\begin{enumerate}
\item \textbf{Question:} Show that there are infinitely many numbers that are divisible by 6 and that can be written as the sum of two prime numbers.
\item \textbf{Solution:} By Euclid's theorem on the infinitude of primes, there are infinitely many primes. By looking at divisibility by two and three, we can see that all primes greater than 3 are either 1 or 5 modulo 6. Therefore, there are \emph{either} infinitely many primes of the form $6k+1$ or infinitely many primes of the form $6k+5$. Let $S$ denote an infinite set of such primes in either case. If the first case holds, then $p + 5$ is divisible by 6 for any $p \in S$. If the second case holds, then $p + 7$ is divisible by 6 for any $p \in S$. In either case, we have found infinitely many numbers that are divisible by 6 and that can be written as the sum of two prime numbers.
\item \textbf{Comment:} The question can also be solved using Dirichlet's theorem on primes in arithmetic progressions, which yields immediately that there are infinitely many primes both of the form $6k+1$ or $6k+5$. However, this is unnecessary and the given solution clearly shows a better understanding of the problem than use of Dirichlet's theorem would. However, we suspected that language models may find the solution using Dirichlet's theorem more natural, as it reduces the problem to something that appears in the training data many times. There are also a number of other elementary approaches to this problem that boil down to showing the infinitude of primes in one of the two arithmetic progressions here of interest; these are also more complicated than the given solution.
\item \textbf{Model \texttt{gpt-4.1}:}
\begin{enumerate}
\item \textbf{Feedback:} Your solution demonstrates some understanding of the structure of primes modulo 6 and attempts to use this to construct sums of two primes divisible by 6. However, there are significant logical gaps and unjustified claims in your argument:

- You correctly note that all primes greater than 3 are congruent to 1 or 5 modulo 6, but your conclusion that there are "either infinitely many primes of the form \(6k+1\) or infinitely many of the form \(6k+5\)" is not justified; in fact, both forms contain infinitely many primes.
- The step where you claim that \(p+5\) or \(p+7\) is divisible by 6 for any \(p\) in your set is not clearly explained or justified, and it is not shown that both terms in the sum are prime for infinitely many cases.
- Most importantly, you do not provide a rigorous argument that there are infinitely many such numbers: you do not construct an explicit infinite family of numbers divisible by 6 that can be written as the sum of two primes, nor do you show that such representations exist for infinitely many numbers.

To improve your solution, you should explicitly construct an infinite family of numbers divisible by 6 that can be written as the sum of two primes, and justify that both terms are prime for infinitely many cases. For example, you could consider numbers of the form \(6n = 3 + (6n-3)\) and argue that since there are infinitely many primes congruent to 3 modulo 6, this gives infinitely many such representations.

In summary, your approach is on the right track in considering the forms of primes modulo 6, but your argument lacks the necessary rigor and explicit construction required for a complete proof. Focus on providing a clear, step-by-step construction and justification to fully answer the question.
\item \textbf{Evaluation:} \texttt{gpt-4.1} completely fails to give useful feedback on this question. Looking at the markscheme constructed by the model in the workflow suggests that the problem can be traced back to it being unable to build a useful markscheme for the question: the markscheme positively mentions approaches that do not work (e.g. writing $6n$ as a sum of $6n-3$ and $3$ and hoping that $6n-3$ is prime). Since the markscheme could be precomputed, it seems likely that the model would do significantly better if it was supplied with a good markscheme in advance; however, the fact that the question is too hard for the model would likely remain problematic.
\end{enumerate}
\item \textbf{Model \texttt{o4-mini}:}
\begin{enumerate}
\item \textbf{Feedback:} Your argument is neat and in fact entirely sufficient.  You avoid any deep analytic input (Dirichlet's theorem) by using only Euclid's theorem and a simple pigeonhole argument on residue-classes mod 6.  Here are a few suggestions to tighten it:

1.  Make explicit that Euclid's theorem gives infinitely many primes \(>3\), and that every such prime is congruent to either 1 or 5 mod 6.  
2.  Then note: since the infinite set of primes \(>3\) must lie in the two classes 1 and 5 mod 6, at least one of these classes is infinite (pigeonhole principle).  
3.  In the case of infinitely many primes $p \equiv 1 \pmod{6}$, the fixed prime $5 \equiv 5 \pmod{6}$ yields $p+5 \equiv 0 \pmod{6}$; analogously, if there are infinitely many $p \equiv 5$, then $p+7 \equiv 0 \pmod{6}$ (and 7 is prime).  Hence in either case you produce infinitely many sums of two primes divisible by 6.  

These steps cover all logical gaps.  Well done on finding this elementary proof.
\item \textbf{Evaluation:} The feedback correctly acknowledges the solution and then makes some suggestions for style improvement. Mentioning the pigeonhole principle explicitly is maybe useful conceptually, but the other suggestions for improvement are nitpicking. Inspection of the markscheme returned by the workflow shows that the model expected a solution using Dirichlet's theorem, so \texttt{o4-mini} shows the ability to fully acknowledge a good solution even if it differs from its own preconceptions. 
\end{enumerate}
\item \textbf{Model \texttt{gpt-5}:}
\begin{enumerate}
\item \textbf{Feedback:} Strengths:
- Correctly identifies that every prime \(>3\) is congruent to 1 or 5 modulo 6 using divisibility by 2 and 3.
- Properly uses Euclid's infinitude of primes to deduce that at least one of the two residue classes (1 mod 6 or 5 mod 6) must be infinite.
- Chooses a fixed prime from the complementary class (5 or 7) and observes that $p + 5$ (for $p \equiv 1 \pmod{6}$) or $p + 7$ (for $p \equiv 5 \pmod{6}$) is divisible by 6, giving representations as sums of two primes.

Minor points to tighten:
- Explicitly justify distinctness: since p ranges over infinitely many distinct primes in S, the sums p + 5 (or p + 7) are all distinct because adding a fixed number is injective.
- State the modular check explicitly (e.g., if $p \equiv 1 \pmod{6}$, then $p + 5 \equiv 0 \pmod{6}$; if $p \equiv 5 \pmod{6}$, then $p + 7 \equiv 0 \pmod{6}$), and note that 5 and 7 are primes, so each sum is indeed a sum of two primes.

Overall, your argument is essentially complete and rigorous; adding a brief note about distinctness (and the quick modulo check) would make it fully watertight.
\item \textbf{Evaluation:} The feedback given by \texttt{gpt-5} here is slightly verbose, but overall good.
\end{enumerate}
\end{enumerate}

\paragraph{Solutions that are strictly speaking correct, but which use overly strong mathematical machinery to solve the problem}
To test the grading of correct solutions that use overpowered machinery, we again used the question from the previous paragraph, but added explicitly as context that the problem is early undergraduate homework material from a course on elementary number theory. We then provided a solution that uses Dirichlet's theorem on primes in arithmetic progressions to solve the problem.

\begin{enumerate}
\item \textbf{Question:} Show that there are infinitely many numbers that are divisible by 6 and that can be written as the sum of two prime numbers. This question is from an early undergraduate course on elementary number theory.
\item \textbf{Solution:} By Dirichlet's theorem on primes in arithmetic progressions, there are infinitely many primes both of the form $6k+1$ and $6k+5$. Let $S_5$ and $S_7$ denote two such infinite sets. Then $p + q$ with $p \in S_5$ and $q \in S_7$ is always divisible by 6, and there are infinitely many such pairs; this yields the desired result.
\item \textbf{Comment:} The solution uses Dirichlet's theorem in a correct way, but using machinery that is beyond the course context. We wanted to test whether the language model would correctly infer that the solution is correct but inadmissible, and if so, what feedback it would produce. The solution additionally contains a minor clerical error in the notation, which is a mathematical issue worth raising in feedback, but which is in our view less significant than the use of overpowered machinery here.
\item \textbf{Model \texttt{gpt-4.1}:}
\begin{enumerate}
\item \textbf{Feedback:} Thank you for your solution. You have correctly identified that Dirichlet's theorem guarantees infinitely many primes in certain arithmetic progressions, and you attempt to use this to construct sums of two primes that are divisible by 6. However, there are some issues in your argument that need attention:

1. **Incorrect Construction:** You claim that for $p \in S_5$ (primes congruent to 5 mod 6) and $q \in S_7$ (presumably you meant primes congruent to 1 mod 6, as 7 mod 6 is 1), the sum $p + q$ is always divisible by 6. However, $5 + 1 \equiv 0 \pmod{6}$, but not all pairs of such primes will sum to a number divisible by 6, since the sum of two odd primes is even, but not necessarily a multiple of 6. Also, your notation $S_7$ is unclear, as there are no primes congruent to 7 mod 6.

2. **Lack of Explicit Infinite Family:** You do not explicitly construct or identify an infinite family of numbers divisible by 6 that can be written as the sum of two primes, nor do you justify why there are infinitely many such numbers. The argument relies on the existence of infinitely many primes in certain congruence classes, but does not show that their sums cover infinitely many multiples of 6.

3. **Missing Justification:** The step from the existence of infinitely many such primes to the conclusion that there are infinitely many sums divisible by 6 is not fully justified. You need to show that for infinitely many $n$, $6n$ can be written as the sum of two primes.

**Suggestions for improvement:**
- Clearly construct an explicit infinite family of numbers divisible by 6 that can be written as the sum of two primes, and justify why both terms are prime for infinitely many such numbers.
- Ensure that your use of congruence classes is accurate and that your argument covers all necessary cases.
- Provide a clear logical link between the existence of primes in certain congruence classes and the existence of infinitely many sums divisible by 6.

Overall, your approach is on the right track by considering arithmetic progressions and sums of primes, but the argument needs to be made more precise and complete to fully answer the question.
\item \textbf{Evaluation:} Inspection of the reasoning chain shows that the model again builds a markscheme very similar to the essentially useless one from the previous question, which is to be expected. The model also wrongly denies that the given construction \emph{works}, despite correctly spelling out the modular arithmetic argument that would show that it does. The rest of the feedback consists of misapplied appeals to rigour and clarity that advance neither. It seems likely that all these symptoms are a result of the broken markscheme construction already mentioned for the previous question. 
\end{enumerate}
\item \textbf{Model \texttt{o4-mini}:}
\begin{enumerate}
\item \textbf{Feedback:} Here is some feedback on your proof:

1. Core idea (good):  
   • You correctly observe that every prime \(>3\) lies in one of the residue classes \(6k+1\) or \(6k+5\).  
   • You then show that if \(p\equiv1\pmod6\) and \(q\equiv5\pmod6\), their sum is a multiple of 6, and hence there are infinitely many such sums once you have infinitely many primes in each class.

2. Main gap (major):  
   • The step "there are infinitely many primes of the form \(6k+1\) and \(6k+5\)" is not elementary; it is essentially Dirichlet's theorem, a deep analytic result.  Since the course is on elementary number theory, one cannot simply invoke this theorem without proof.  

3. Notation remark (minor):  
   • You denote your two infinite prime-sets by \(S_5\) and \(S_7\), but you really mean ``primes $\equiv 5 \pmod{6}$'' and ``primes $\equiv 1 \pmod{6}$''.  It would be clearer to call them, say, \(S_{5}\) and \(S_{1}\).  

4. How to strengthen:  
   • Either supply (or reference) a proof that there are infinitely many primes in each of the classes \(1\pmod6\) and \(5\pmod6\), or else reframe the argument so it does not hinge on that unproved fact.  In an elementary-theory course one normally avoids Dirichlet's theorem unless it has already been covered.  

Overall you have the correct construction in principle, but the reliance on a heavy unproved theorem is too large a gap in this setting.
\item \textbf{Evaluation:} The model correctly criticises the notation error and the use of Dirichlet's theorem given the course context. Feedback is lengthy, but mathematically correct; however, the markscheme remains largely generic, which suggests that the model itself was not able to develop an elementary solution. The feedback could be shortened, but is overall nonetheless good.
\end{enumerate}
\item \textbf{Model \texttt{gpt-5}:}
\begin{enumerate}
\item \textbf{Feedback:} Feedback:

- What you did well:
  - Correctly identified the relevant residue classes: primes \(>3\) are $6k \pm 1$, and $1 + 5 \equiv 0 \pmod{6}$, so such pairs sum to a multiple of 6.
  - The overall strategy---pairing a prime $\equiv 1 \pmod{6}$ with a prime $\equiv 5 \pmod{6}$---is appropriate.

- Main issues to address:
  - You invoke Dirichlet's theorem without proof. For this course and problem, relying on such a deep result is not acceptable; an elementary argument is expected.
  - Even if Dirichlet were allowed, you should explicitly ensure you obtain infinitely many distinct multiples of 6, not merely infinitely many pairs. As written, "there are infinitely many such pairs" does not by itself rule out repeated sums.

- How to strengthen your solution within the course scope:
  - Replace the appeal to Dirichlet with an elementary Euclid-style argument proving that there are infinitely many primes in at least one of the classes $6k+1$ or $6k+5$. A standard route is: assume finitely many primes $\equiv 5 \pmod{6}$, form $N = 6\prod r_i - 1$, show $2 \not\mid N$ and $3 \not\mid N$, and that $N$ has a new prime divisor $q \equiv 5 \pmod{6}$.
  - Having established such an infinite set (say, primes $q \equiv 5 \pmod{6}$), fix a single prime $r \equiv 1 \pmod{6}$, for example $r = 7$, and note that for each $q$, the sum $q + r$ is divisible by 6 and these sums are all distinct because $q$ varies. Briefly justify that $(q + r)/6$ is a natural number.

- Minor note:
  - There is a notation slip: you wrote $S_5$ and $S_7$, but you appear to mean the sets of primes $\equiv 5$ and $\equiv 1 \pmod{6}$.

Addressing these points will make your argument rigorous and self-contained at the intended level.
\item \textbf{Evaluation:} The feedback addresses all the main points and points out a viable elementary strategy to fix the proof. The only drawback is that the model does not find or hint at the more elegant strategy of not deciding which of the two residue classes is infinite. The notational slip-up is also found.
\end{enumerate}
\end{enumerate}

\paragraph{Advanced questions}
To test how the model would deal with a question that goes beyond its knowledge horizon, we designed a question that could plausibly be given in an early graduate-level course on cryptography, or possibly in an undergraduate course if the students were taught exactly the right background material. We hoped that the underlying language model would fail to solve the question, and were interested how it would then deal with checking an erroneous but convincing-soundingsolution.

\begin{enumerate}
\item \textbf{Question:} Let $N = pq$ be an RSA modulus, where $p$ and $q$ are distinct primes chosen uniformly at random from the set of all primes of size 1024 bits. Assume that the public key is $(N, e)$, with $e = 65537$, and the private exponent is $d$. Eve wants to learn as much as possible about the private key $d$, but unfortunately her quantum computer is broken and she cannot factor $N$. However, she does know the SHA-256 hash value of the first 1000 binary digits of $d$, when these binary digits are interpreted as a binary number represented as a string. Determine how many bits of $d$ she can learn from this information.
\item \textbf{Solution:} SHA-256 is a one-way hash function, and therefore Eve's knowledge of the hash value of the first 1000 binary digits of $d$ does not reveal any practically useful information about $d$. Eve does have some information about $d$ that is given by $d$ obeying $d \equiv e^{-1} \mod \phi(N)$, but this is not enough to learn any bits of $d$ without knowing $\phi(N)$; and knowing $\phi(N)$ would allow her to probabilistically factor $N$. Therefore, unless $N$ is of some special form, Eve cannot learn any bit-values of $d$ from this information.
\item \textbf{Comment:} The solution is wrong on multiple levels. A trick due to Boneh, Durfee and Frankel \cite{boneh1998attack} can be used to constrain the top half of the bits of $d$ to a small set of possibilities, and the hash value disambiguates between them. She can therefore learn roughly the top half of the bits of $d$ from this information. Additionally, some less fundamental confusions are present in the solution; for instance, knowledge of $\phi(N)$ allows Eve to deterministically factor $N$, not just probabilistically. Good feedback should make it clear to the student where they go wrong most fundamentally, ideally without revealing the solution.
\item \textbf{Model \texttt{gpt-4.1}:}
\begin{enumerate}
\item \textbf{Feedback:} Your solution correctly identifies that SHA-256 is a one-way hash function and that its preimage resistance prevents Eve from learning any useful information about the first 1000 bits of $d$ from the hash value alone. You also note that, without the ability to factor $N$ or compute $\phi(N)$, Eve cannot exploit the RSA structure to deduce $d$. 

However, your answer could be improved by explicitly stating that, due to the strong preimage resistance of SHA-256, Eve cannot feasibly recover any of the 1000 bits of $d$ or even partial information about them, except with negligible probability. It would also be helpful to state clearly that the number of bits Eve can learn is zero, aside from what is trivially implied by the hash value itself (i.e., that the prefix of $d$ hashes to the given value).

Overall, your reasoning is sound and your answer is essentially correct, but a more explicit reference to the preimage resistance property and a direct statement about the number of bits learned would make your argument more rigorous and complete.
\item \textbf{Evaluation:} The model accepts the answer as correct although it is wrong. Its suggestions for improvement are not helpful. Looking at the reasoning chain shows that the main reason for this is that the model itself converged towards a very similar answer to the one given as the student submission. Likely, again a better precomputed markscheme would improve the situation significantly.
\end{enumerate}
\item \textbf{Model \texttt{o4-mini}:}
\begin{enumerate}
\item \textbf{Feedback:} Your answer correctly recalls that SHA-256 is one-way and that without factoring N one cannot pin down d exactly. However, it stops short of the key quantitative step required by the question: you must show how many bits of d can be learned in principle from a 256-bit hash of its first 1000 bits.

1. Quantify the leakage.  A 256-bit digest can convey at most 256 bits of information about its 1000-bit preimage.  Equivalently, if D is the 1000-bit prefix of d, then  
   $I(D; \text{SHA256}(D)) \leq 256$,  
   so $H(D | \text{SHA256}(D)) \geq 1000 - 256 = 744$ bits remain uncertain.  Thus Eve can learn at most 256 bits of D.

2. Tie to preimage resistance.  One-wayness alone only tells you "inverting is hard," but does not quantify how much comes out of the hash.  By invoking the bound on mutual information (or Shannon entropy) you show rigorously that a 256-bit hash cannot reveal more than 256 bits of its 1000-bit input.

3. Conclude clearly.  Therefore---even with an unbroken hash oracle---Eve can learn at most 256 bits of the first 1000 bits of d.  All remaining bits still carry at least 744 bits of uncertainty.

In future, whenever a hashing question asks "how many bits does this leak?" be sure to back your answer with an explicit entropy or information-theoretic inequality relating preimage size to digest length.
\item \textbf{Evaluation:} The feedback incorrectly says that the solution is correct up to not providing quantitative estimates of leakage. Inspection of the markscheme suggests that the model expects a solution along the lines provided, but with the quantitative arguments it asks for. It does not seem unlikely that a better markscheme step would improve results, given that the model applies the underlying concepts in a mostly correct way but lacks one key idea. It fails to mention the minor erroneous claim made in the solution that knowledge of $\phi(N)$ only allows \emph{probabilistic} factoring.
\end{enumerate}
\item \textbf{Model \texttt{gpt-5}:}
\begin{enumerate}
\item \textbf{Feedback:} Feedback

What you did well
- You reached the correct conclusion: from the SHA-256 of the first 1000 bits of d, Eve cannot determine any bits of d.
- You appealed to the right hash property (one-wayness/preimage resistance) and correctly noted that RSA relations like $ed \equiv 1 \pmod{\phi(N)}$ don't help without $\phi(N)$.
- You correctly observed that knowing $\phi(N)$ would immediately reveal the factors (hence d), so the hash does not assist with factoring.

How to strengthen your argument
- Make the computational vs. information-theoretic point explicit. Under standard assumptions (preimage resistance / random oracle behavior), knowing a 256-bit digest of a 1000-bit prefix leaves about $2^{(1000-256)} \approx 2^{744}$ candidate prefixes consistent with the hash. Thus no specific bit of the prefix (and hence no bit of d) is fixed or predictable with non-negligible advantage over 1/2.
- Quantify the candidate count/entropy: the digest ``removes'' about 256 bits of entropy from the 1000-bit prefix but does not determine any particular bit.
- Be precise about the RSA relation: write $ed \equiv 1 \pmod{\phi(N)}$ (or $\lambda(N)$), and note that this gives no leverage to connect the hash to d.
- Minor correction: given $\phi(N)$, factoring $N$ (and thus finding d) is deterministic and efficient, not merely ``probabilistic.''

Suggested revision of key justification (for clarity only)
- ``Under preimage resistance/random oracle assumptions for SHA-256, the set of 1000-bit strings mapping to a given 256-bit hash has size about $2^{744}$. Therefore, conditioning on the observed hash does not determine any particular bit of the 1000-bit prefix, and no bit of d is learnable with non-negligible advantage. Hence, Eve learns 0 bits of d.''
\item \textbf{Evaluation:} The model makes the same error as the other models and calls the solution correct because it is essentially the solution that it itself expects. The proposed changes do not improve the solution in any way.
\end{enumerate}
\end{enumerate}
\paragraph{Multi-lingual support}
To test this, we used the question from the previous paragraph, but this time provided a \emph{correct} solution in German.
\begin{enumerate}
\item \textbf{Question:} Let $N = pq$ be an RSA modulus, where $p$ and $q$ are distinct safe primes of size 1024 bits each. Assume that the public key is $(N, e)$, with $e = 65537$, and the private exponent is $d$. Eve wants to learn as much as possible about the private key $d$, but unfortunately her quantum computer is broken and she cannot factor $N$. However, she does know the SHA-256 hash value of the first 1000 binary digits of $d$, when these binary digits are interpreted as a binary number represented as a string. Determine how many bits of $d$ she can learn from this information.
\item \textbf{Solution:} Wir haben $\varphi(N) = (p-1)(q-1) = N - p - q + 1$, also $\varphi(N) \approx N$, mit einem Fehler in der Größenordnung $\sqrt(N)$ (da $p$ und $q$ etwa gleichgroß sind). Außerdem ist $ed = 1 + k\varphi(N)$ für ein $k \in \mathbb{N}$ mit $k < e$. Daraus folgt $d = (1 + k\varphi(N))/e$ für dieses $k$. Die führenden Bits von $d$ können nun über die bereits erwähnte Approximation $\varphi(N) \approx N$ ermittelt werden, da damit $d \approx (1 + kN)/e$ ist. Hierbei gibt es noch $\approx 2^{16}$ Möglichkeiten für $k$. Eve kann sämtliche Möglichkeiten durchprobieren und dasjenige $k$ finden, das mit dem gegebenen Hash-Wert übereinstimmt. Auf diese Weise lernt sie etwa die ersten $1024$ Bits von $d$.
\item \textbf{Comment:} The solution is largely correct, but is given in German. It contains a minor overclaim in that the approximation error may cause carry errors a short way up from the 1024-bit mark, which Eve cannot correct; it also fails to note that Eve can recover the lowest bit of $d$ by noting that $d$ must be odd. We wanted to see whether our system would give suitable feedback (ideally noting that the solution in principle correct and very briefly noting the carry issue), and what language it would choose to give it in; and also, whether the system would recognise that the solution is correct.
\item \textbf{Model \texttt{gpt-4.1}:}
\begin{enumerate}
\item \textbf{Feedback:} Your solution demonstrates a good understanding of the structure of RSA and the relationship between $d$, $e$, and $\varphi(N)$. However, your argument does not address the core cryptographic aspect of the question: the security properties of the SHA-256 hash function and what information (if any) is leaked by knowing the hash of the first 1000 bits of $d$.

Specifically, your solution suggests that Eve can try all possible values of $k$ (with $k < e$) and, using the approximation $\varphi(N) \approx N$, compute candidate values for $d$ and check which one matches the given hash. However, this approach overlooks two crucial points:

1. **Preimage Resistance of SHA-256:** SHA-256 is designed to make it computationally infeasible to recover the input (the first 1000 bits of $d$) from its hash. Even if Eve can generate candidate values for $d$, she cannot efficiently check all $2^{1000}$ possible 1000-bit prefixes, nor can she invert the hash to recover the prefix.

2. **Number of Candidates:** While there are only about $2^{16}$ possible values for $k$, the corresponding values of $d$ are extremely close to each other and differ in the lower bits. The first 1000 bits of $d$ are not uniquely determined by $k$ without knowledge of $p$ and $q$, and the hash does not help Eve distinguish between them due to the preimage resistance of SHA-256.

In summary, under standard cryptographic assumptions, knowing the SHA-256 hash of the first 1000 bits of $d$ does not allow Eve to learn any significant information about $d$. The number of bits she can learn is essentially zero, except with negligible probability.

To improve your answer, focus on the cryptographic properties of SHA-256 and explain why these prevent Eve from learning the bits of $d$ from the hash, regardless of her knowledge of the RSA structure.
\item \textbf{Evaluation:} The model again follows the broken markscheme to the letter, and therefore criticises the differences between the proposed solution and what the markscheme demands the student do, without realising that the student solution is in fact largely correct. 
\end{enumerate}
\item \textbf{Model \texttt{o4-mini}:}
\begin{enumerate}
\item \textbf{Feedback:} Your write-up shows you've thought about relating $ed \equiv 1 \pmod{\phi(N)}$ and the small range of $k<e$, but the core of the question is that a secure hash of 1000 bits simply cannot be inverted or ``matched'' to any candidate d-prefix without a truly astronomical search---so in fact Eve learns 0 bits. Here are the main issues with your argument:

1.   Misplaced reliance on $\phi(N) \approx N$.  
     You observe $\phi(N)=N-(p+q)+1$ and note that $p+q \approx 2\sqrt{N}$, so $\phi(N)$ and $N$ differ by roughly $\frac{1}{2}$ bit-length worth of ``noise'' (on the order of $2^{1024}$). That error far outweighs any gain from knowing $k<2^{16}$, so you cannot pin down the first $\approx 1024$ bits of d to a unique value. Any small error in $\phi(N)$ propagates directly into $d \approx (1+kN)/e$, making those ``leading bits'' completely undetermined.

2.   Ignoring hash preimage resistance.  
     SHA-256 is designed so that even if you can efficiently enumerate all $2^{16}$ values of k, you would still need to invert a 256-bit hash on a 1000-bit input. Brute-forcing over all possible 1000-bit prefixes remains infeasible ($2^{1000}$ possibilities), so the hash gives essentially no usable information about those bits of d.

3.   Conclusion contradicts cryptographic assumptions.  
     A correct solution invokes that d has about 2048 bits of entropy, that SHA-256 resists preimage attacks up to $\approx 2^{256}$ work, and that knowing the hash of its first 1000 bits leaks no information about those bits. Hence Eve cannot recover any bits of d from the hash alone.

To improve, you should:  
•   Explicitly state the preimage-resistance assumption for SHA-256 and quantify why inverting the hash is infeasible, even if k is small.  
•   Show that approximating $\phi(N)$ by $N$ leaves an uncertainty far larger than the 1000 bits you hope to learn.  
•   Conclude rigorously that the number of bits learned is zero, under standard cryptographic assumptions.
\item \textbf{Evaluation:} The model wrongly ``corrects'' a perfectly good solution. Again, this could be due to the markscheme, which is very similar to the one built for the previous test question and expects a wrong solution. 
\end{enumerate}
\item \textbf{Model \texttt{gpt-5}:}
\begin{enumerate}
\item \textbf{Feedback:} Thank you for the careful reasoning. However, your conclusion and the way you use the hash value are not in line with the cryptographic setting of the problem.

Key issues:
- You never justify, from a computational-security standpoint, that knowing $H = \text{SHA-256}$ of the first 1000 bits of d lets Eve recover any bit of that prefix. Under standard assumptions (preimage and second-preimage resistance of SHA-256), given only $H$ of a 1000-bit string, no efficient algorithm can compute any specific bit (or any nontrivial predicate) of that string with better than negligible advantage. Thus, the number of bits of d learned is 0.
- The RSA algebra you use ($ed = 1 + k\phi(N)$ with $k < e$) and the facts that p, q are safe primes and $e = 65537$ do not give computational leverage to extract bits of d from $H$. They are irrelevant to what can be inferred from a hash of a prefix.
- The approximation $\phi(N) \approx N$ is not a valid bridge from $H$ to the true prefix. The difference $N - \phi(N) = p + q - 1$ is unknown and large (about $2^{1024}$). While you estimate that $d \approx (1 + kN)/e$ and propose brute-forcing k, you do not establish a cryptographically sound guarantee that the first 1000 bits of the true d equal the first 1000 bits of your approximant for the correct k, nor that $H$ can be used to validate that equality without effectively inverting the hash on a 1000-bit input. Invoking an approximation does not circumvent the one-wayness of SHA-256.

What would earn full credit:
- State clearly that, under the preimage (and second-preimage) resistance of SHA-256, from $H$ alone one cannot compute any particular bit (or any nontrivial predicate) of the 1000-bit prefix with non-negligible advantage; hence Eve learns 0 bits of d. Also note explicitly that the safe-prime structure of $N$ and the choice $e = 65537$ are irrelevant to this conclusion.
\item \textbf{Evaluation:} Again, like the other models, the model does not accept the solution clearly in part because it is not the one expected by the markscheme. It is surprising that the model calls the reasoning ``careful'', but rejects it wholesale without giving coherent reasoning.
\end{enumerate}
\end{enumerate}

\paragraph{Unconventional questions}
Here we designed a question that is absolutely undergraduate-level, and in fact easy, but slightly counter-intuitive. Our question takes the form of a riddle that can be solved by application of some simple probability theory. We supplied a solution that is subtly but fundamentally wrong. The question was then to see whether our system would correctly identify the mistake.

\begin{enumerate}
\item \textbf{Question:} In a galaxy far, far away, there is a planet which has been settled by humans. The humans have built a very advanced civilisation, but their numbers are fairly low; robots and AI do most of the work. There are about 1000 humans in total on the planet, most of them living in one village. Recently, they have come up with a cure for aging, so nobody dies anymore. Now, if nobody ever dies, this presents in the long run some resource problems that even a post-scarcity society cannot solve. In order to prevent this, they have passed a law that says that no person may have more than two children. Assuming a random 1:1 sex ratio at birth, and that nobody ever dies, and that generation time is about 30 years, and that a Malthusian catastrophe occurs if the planet hits more than $10^{10}$ inhabitants, what will their population roughly be a billion years from now? You may assume that an overpopulation catastrophe leads to a decimation by a factor of 10 within a short time span, but that normal conditions resume shortly thereafter because the robots rebuild. Calculate also the long-term trajectory beyond that point in time.
\item \textbf{Solution:} Since nobody ever dies, and since everyone has at most two children, and since the gender ratio is 1:1, every generation will have exactly 1000 births (since each child has two parents). Therefore, initial growth is 1000 births every 30 years, which is a growth rate of $1000 / 30 \approx 33.33$ births per year. After a billion years, the population will be roughly $1000 \times 10^9 / 30 \approx 3.33 \times 10^{10}$ inhabitants. This is clearly more than $10^{10}$, so a Malthusian catastrophe will occur. The first such catastrophe will occur after $10^{10} / 33.33 \approx 300$ million years. It will reduce the population to about $10^9$ people, the vast majority of whom will already have had two children; assuming deaths in the catastrophic state are random, the first Malthusian catastrophe reduces future births to $\approx 3.33$ per year, because it reduces the remaining legally breeding population by a factor of 10. A 700-million year climb follows which takes the final population to $\approx 10^9 + 3.33 \cdot 7 \cdot 10^8 \approx 2.33 \cdot 10^9$. In the long term, catastrophe states repeat at ever-increasing intervals, each reduces the breeding population by 90 percent, and the population finally hits a steady state at roughly one billion inhabitants.
\item \textbf{Comment:} The solution is designed to show a failure mode that is plausible for human reasoning. It fails to properly model the growth process as a stochastic process. The limiting factor for growth is the size of the male or female population who have had less than two children as yet, and this quantity should be modelled by a random walk where each birth increases either males or females by one, and deducts one from the other gender. This random walk will eventually hit the boundary where one gender has no members with remaining reproductive capacity, and the population is subsequently eternally stable. 
\item \textbf{Model \texttt{gpt-4.1}:}
\begin{enumerate}
\item \textbf{Feedback:} Your solution demonstrates a good attempt to model the population dynamics under the given constraints, and you correctly identify that the two-child policy with no deaths should, in principle, lead to a stable population in the long term. You also recognize the effect of the Malthusian catastrophe and describe the resulting sawtooth pattern in population size.

However, there are some important issues in your reasoning:

1. **Population Growth Under Two-Child Policy:**  
   The key point is that with a strict two-child policy and no deaths, the population should stabilize, not grow. In your solution, you assume that every person can have two children, leading to a constant number of births per generation, but you do not account for the fact that, with no deaths, the population accumulates over time. However, the two-child policy is typically interpreted as a per-couple limit, which means that each couple replaces themselves, and the population should stabilize at twice the number of couples (i.e., the initial population, if everyone has two children and there are no deaths). Your calculation of a linear increase in population over time is not consistent with this logic.

2. **Calculation of Population After a Billion Years:**  
   You estimate the population after a billion years by multiplying the number of births per year by the number of years, which assumes that all births add to the population without any constraint. This ignores the stabilizing effect of the two-child policy and overestimates the population growth.

3. **Modeling the Catastrophe:**  
   You correctly note that when the population hits $10^{10}$, a catastrophe reduces it by a factor of 10. However, your subsequent calculation of the post-catastrophe growth rate assumes that the number of people able to have children is reduced proportionally, but you do not justify how the breeding structure is maintained or restored after the catastrophe. In reality, after such a reduction, the age structure and the number of people able to have children may not allow for immediate resumption of the previous growth rate.

4. **Long-Term Trajectory:**  
   You correctly describe that the population will oscillate between $10^{10}$ and $10^9$ due to repeated catastrophes, but the underlying assumption of continued linear growth between catastrophes is not justified under a strict two-child policy with no deaths.

**Suggestions for Improvement:**
- Carefully consider the implications of the two-child policy: with no deaths, the population should stabilize, not grow indefinitely.
- When modeling the effect of the catastrophe, discuss how the age and reproductive structure of the population is affected, and whether the population can resume growth at the same rate.
- Explicitly justify each step in your reasoning, especially when making assumptions about how the population changes after each catastrophe.

Overall, your solution shows a good understanding of the scenario and the key mechanisms at play, but it contains a significant logical error regarding the long-term population growth under the two-child policy. Focusing on the stabilizing effect of the policy and clarifying your assumptions about the population structure will strengthen your argument.
\item \textbf{Evaluation:} The model suggests that a two-child policy automatically leads to stable population size, and does not coherently integrate the immortality assumption of the problem. Again, this can be traced back to a poor markscheme which does not present a coherent solution and makes logically incompatible demands on the solution (in particular, the student is required to both predict a stable population size \emph{and} account for Malthusian catastrophies). It seems possible that a better markscheme would improve results, but the level of misunderstanding the problem displayed here by the model is very significant and makes it seem doubtful that simply fixing the markscheme manually would resolve possible \emph{other} failure cases than the one here tested.
\end{enumerate}
\item \textbf{Model \texttt{o4-mini}:}
\begin{enumerate}
\item \textbf{Feedback:} Your write-up shows initiative in trying to track births, catastrophes, and long-term behaviour, but there is a fundamental flaw in your growth model that undercuts the rest of the argument.  In particular:

1.  Mis-modelling the birth rate as a constant "1000 births every generation" neglects the obvious fact that if no one ever dies and each couple has two children, then the number of births in a generation grows with the existing population.  In fact each pair of parents produces two offspring, so the total births per generation is proportional to the number of parents (half the population), and hence the population doubles each generation rather than increases by a fixed amount.  This is why the process is exponential, not linear.

2.  Because of that, the timing of reaching $10^{10}$ inhabitants (and thus the timing of catastrophes) is set by an exponential recurrence, not by dividing a billion years by a constant annual birth rate.  You would need to write something like  
   $P_{n+1} = 2 \cdot P_n$ until $P_n \geq 10^{10}$,  
   then P resets to $P_n/10$,  
   and so on.  
   From this you can compute the number of generations between disasters $\approx \log_2(10)$, and how many such cycles fit into $10^9$ years (using 30-year generations).  

3.  Once you have the correct recurrence, you can count how many catastrophes occur in $10^9$ years (it ends up being on the order of a few tens, not hundreds of millions of years), and show that the population oscillates around $10^{10}$ rather than drifting linearly to $10^9$ as you found.

Next steps to improve:

--  Derive the per-generation multiplier carefully (why each generation roughly doubles).  
--  Incorporate the catastrophe threshold piecewise into your recurrence.  
--  Estimate the cycle length in generations and the total number of cycles over $10^9$ years.  
--  Show that the process settles into a bounded oscillation near $10^{10}$, rather than a linear climb to $10^9$.
\item \textbf{Evaluation:} The proposed ``improvements'' would make the solution substantially worse. In particular, the proposed student solution at least coherently explores the behaviour of the target function in a deterministic growth model. With the changes proposed by the model, it is not clear any more that there is any coherent model of the process still being analysed.
\end{enumerate}
\item \textbf{Model \texttt{gpt-5}:}
\begin{enumerate}
\item \textbf{Feedback:} Feedback

What you did well
- You correctly recognized that the two-children-per-person rule with a 1:1 sex ratio implies linear (not exponential) growth with a constant number of births per 30-year generation.
- Your estimate of the first overpopulation hit at about $3 \times 10^8$ years is on target.
- You correctly noted that a factor-10 decimation reduces the subsequent birth rate by the same factor, so post-crash growth is much slower and the intervals between catastrophes increase.

Points to improve/clarify
- Justification of the birth rate B: ``each child has two parents'' does not imply 1000 births per generation. A sound argument is: with about 1000 people initially, roughly 500 are women; each woman is allowed at most 2 children, so one 30-year cohort can produce at most about $500 \times 2 \approx 1000$ births. This is the replacement-level cohort size that drives linear growth in this model.
- 1 Gyr population arithmetic: after the first crash (at $\sim 0.3$ Gyr) to $10^9$, you have 0.7 Gyr of regrowth at $\sim 3.33$ per year. That adds about $3.33 \times 7 \times 10^8 \approx 2.33 \times 10^9$ people, so the total is about $10^9 + 2.33 \times 10^9 \approx 3.33 \times 10^9$, not $2.33 \times 10^9$. This looks like a simple addition slip.
- Long-run behavior: it does not converge to a steady state at $10^9$. The process is piecewise linear growth punctuated by decimations back to $10^9$, with each decimation also reducing the slope by a factor of 10. Consequently, the intervals between catastrophes grow by a factor of about 10 each time (e.g., next crash around 2.7 Gyr, then around 27 Gyr, etc.), but the cycle does not settle to a fixed population.

Suggestions
- Make the discrete-generation model explicit: $N_{k+1} = N_k + B$ pre-crash with $B \approx 1000$ per 30 years; after each decimation, $N$ is divided by 10 and $B$ is divided by 10, giving a clear piecewise-linear trajectory.
- When reporting headline numbers (like at 1 Gyr), double-check the final addition to avoid losing the base level after the crash.
\item \textbf{Evaluation:} The model gives some practical guidance on the justification and implementation details of the simple deterministic model used by the proposed solution. It completely misses the fact, however, that this deterministic model is wrong for the problem at hand. It incorrectly criticises the proposed solution also for predicting an eventual steady state, because it does not appreciate properly that even the deterministic process is discrete.
\end{enumerate}
\end{enumerate}

\subsection{Lessons Learned and Practical Guidance}\label{sec:lessons}

This section gives a summary of the lessons we learned from both our experiments and our work deploying the tutoring application in \lambdafeedback (a pre-existing feedback generation application at \imperial). The results we discuss here are mostly qualitative in nature; quantitative results are reported in Subsection \ref{sec:quantitative_results} and Appendix \ref{app:detailed_correlation_results}.

\paragraph{Prompting and workflow}

In our testing, a modular workflow was essential for obtaining acceptable feedback, especially when working with \emph{non-reasoning} LLMs. When asking the model directly to generate feedback to a given question and solution, output was frequently too verbose and/or at least partially wrong. Non-reasoning models in this setting are forced to simultaneously figure out the solution and the feedback, and do so without revealing the solution via their chain-of-thought to the student. Nonetheless, if a separate grading step is added, non-reasoning models are able to produce good grade-correlations with direct feedback (see Section \ref{sec:quantitative_results}) if a relatively strong model is used for the grading step. However, this could plausibly be due to the grading model ignoring the feedback and grading based on the question and solution alone. In principle, this could be easily tested by running the grading step in isolation while randomly permuting the feedback and the question/solution pairs to be graded; if the grade-correlation is still high, this would very strongly suggest that the feedback is being ignored. We leave such tests for future work.

Decomposing the task also opens up the possibility of performing parts of the task (for instance, generation of model solutions or detailed grading rubrics) in a one-time precomputation step, which saves cost and allows the use of the most powerful models available for this part. This is very relevant for deployment settings where the use case is not to process \emph{arbitrary} questions, but rather to give feedback on questions assigned to the student by a teacher. Precomputation of such steps also has the advantage that human teachers can directly edit the precomputed steps to improve accuracy or pedagogical quality or to otherwise tune the system to produce better feedback. More generally, task decomposition greatly simplifies debugging of the tutoring system, because each step can be independently audited.


We also experimented with varying the model's \emph{persona}, by which we mean the role the model is instructed to play when performing the various steps of a given workflow. Adopting a \emph{teaching-assistant} persona—clear, supportive, and non-hedged—consistently produced more actionable feedback for first-year proofs than either neutral or overly cautious voices. Experiments with making the LLM model (in this case, \texttt{gpt-4o}) act as a specific expert in mathematics (in this case, Terence Tao) did not yield noticeably better results either along our quantitative metrics or in subjective assessment than the teaching-assistant persona.

For non-reasoning models, one parameter that is a natural target for tuning is the \emph{temperature} used in the inference process.\footnote{For reasoning models, sampling temperature is commonly not exposed by the model API.} Roughly speaking, increasing the temperature flattens the distribution of tokens selected for output, while lowering it makes the model more deterministic. The models offered by OpenAI do not seem to be deterministic at zero temperature. For steps that can be run in precomputation, it is very plausible that results can be improved and sanity-checked by running the step multiple times and storing a majority vote or flagging significant disagreements for human review. However, in our experiments, we did not perform such best-of-N runs.



 \paragraph{Data augmentation} We attempted to augment the dataset with more answers generated automatically to match a given target grade. This yielded very unsatisfactory results, as the models we tried seemed unable to generate wrong answers with natural mistakes on command. It would certainly have been possible to make language models generate wrong answers to the given questions, for instance by choosing a sufficiently weak language model, but such answers would likely not have produced natural human error patterns and it would have been impossible in advance to predict the severity of the mistakes.

\paragraph{Failure modes and mitigations}
We now discuss some common failure modes and mitigations that we found useful in our testing: 


\begin{itemize}
\item Feedback is often still verbose. This is especially true for feedback on correct solutions. It is possible to mitigate this problem by explicitly limiting the length of the feedback, but this needs to be balanced against the need to provide the student with all of the information they need to learn from the feedback; in our opinion, neither the human feedback nor our better LLM-generated workflows hit the mark on this, with the human feedback tending towards too much concision and the LLM-generated feedback being too verbose. Further experimentation and user testing will be needed to find a good balance.
\item Feedback on correct solutions is often padded with suggestions for stylistic improvements. Qualitatively, we have seen this less with reasoning models than with non-reasoning models, which for the question sets considered in this work is one of the few tangible benefits of using reasoning models instead of sufficiently capable non-reasoning models.
\item When the tutoring application is integrated into a pre-existing e-learning platform, compatibility issues can arise. For instance, traditional intelligent tutoring systems rely on dedicated algorithms to check student answers. These dedicated algorithms cannot process the question in the natural-language format that the student is shown, and indeed frequently do not even have to know what the question is (instead, it is sufficient for them to know what the expected \emph{answer} is). A consequence for systems such as ours is that the platform will have separate storage for reference solutions and questions, and a connection between the two cannot be expected to be automatic. This can easily lead to misconfigurations and problems with the human review of question setup on the platform. Automated setup scripts are a mitigation, but conversion from natural-language mathematical problem sheets to the format required for storage on \lambdafeedback{} for instance proved nontrivial to automate fully. In principle, dedicated online platforms for ITS systems that process problems posed in natural language should probably make the question setup directly visible to the tutoring system.
\item Supplying context from the course about notations, conventions, and allowed results is very important to allow the system to correctly assess student answers. The underlying models have otherwise no way of knowing these course-specific details. It should be possible to automatically extract this context from lecture notes or to guess plausible defaults from the automatic processing of problem sheets, but human review of such assumed defaults is necessary and the teaching platform has to support it.
\item Deployments must balance processing cost with accuracy of generated feedback. For mathematically more complex questions, it is likely beneficial to use reasoning models for the feedback generation step, even though the present work does not show any benefit from doing so. To allow for this, deployments should allow model defaults to be overridden on a per-question or even per-workflow-step basis. Additionally, automated processing of user feedback could be used to escalate to stronger models if the user either requests it or the system detects that they are having trouble with the feedback.
\item For questions that overwhelm the system's mathematical reasoning abilities, our stress testing showed instances of very poor output. We expect that some of our stress testing questions would also be wrongly graded by mathematically competent but not specifically prepared human graders. Still, in settings with harder questions, our stress testing shows that selecting underlying language models with strong reasoning capabilities is important. Additionally, when parts of a workflow can be run without knowing the student submission and the results of those workflow steps can be reused across many submissions, it makes sense to use the strongest available models to run these steps and/or to review or modify these steps using human expertise.
\end{itemize}

Overall, we found that the system presented is quite robust in the sense that a wide range of errors in the submitted solutions are correctly identified and the feedback is helpful. Nonetheless, there is certainly still room for improvement, which we hope to achieve in future work.

\subsection{Open Questions and Future Work}\label{sec:open_questions_future_work}

The current paper presented work on the construction and evaluation of a modular feedback generation system for undergraduate university-level mathematics. Unlike other work we are aware of, our work focused to a large extent on the evaluation of feedback correctness. We hope to have shown that the system as presented is already a useful tool for feedback generation and related tasks (such as proofreading of exemplary solutions) for higher mathematics education. It should open up various possibilities for follow-up work. We will briefly discuss some of these in this section.

\paragraph{Integrating Large Language Models with Traditional Intelligent Tutoring Systems} In the context of deploying our mathematics tutor on \lambdafeedback{}, we did a significant amount of work to automatically convert problem sheets presented in their traditional format to the format required by the platform and by our system. This work naturally lends itself to the follow-up question of whether similar automations could be made available to other types of tutoring systems. Essentially, this would involve developing an intelligent interface that allows a teacher to give their online platform their problem sheets, and have applicable questions converted to the format required by an e-learning platform that uses specialised algorithms to check student answers.

\paragraph{Integrating Formal Theorem Provers or Other Mathematical Tools} It could be interesting to develop interactive teaching environments similar to ours that support natural-language mathematical problem solving and interaction with the teaching system, but which connect to computer algebra systems or formal theorem prover backends to perform complete verification of the student's solution. 

In the setting where an LLM would be working in tandem with Lean or another proof assistant, it would be expected to handle on the one hand autoformalisation (and this has been studied for instance in LeanTutor \cite{patel2025leantutor}), but could also comment both on logical gaps in the student's argument that become apparent after autoformalisation, or on high-level strategy and presentation issues. We expect that this could be very useful in some mathematics teaching contexts, for instance where having a computational sandbox to run experiments in is beneficial.

\paragraph{Better Evaluation of Feedback Quality} The methods used for the evaluation of feedback quality in the present work are relatively limited. The end goal of teaching is to teach mathematics to students who do not yet know it, and of course at the other end of the evaluation methods spectrum, practical longitudinal studies of student learning are the gold standard, but they are also very expensive to run, need ethics review, and are therefore often limited in scope. It would be very interesting to develop automated proxies for pedagogical quality of feedback. For instance, one idea could be to use the in-context learning capabilities of language models to simulate human learners solving the same questions as the students in the dataset, and to then compare the solving ability of a particular model on new questions before and after receiving feedback on questions attempted.

\paragraph{More Intuitive Workflow Configuration} The workflow configuration interface used in our prototype requires only knowledge of a natural language (with the tested language being English, but other widely spoken languages expected to work as well), and a basic understanding of how to use a JSON file. However, in many settings, this still leaves some room for improvement. Ideally, teaching staff should be able to set and experiment with new workflows just by \emph{drawing} a workflow graph similar to the one shown in Figure~\ref{fig:pipeline_architecture}. We expect that this should be achievable; possibly even with something as simple as an LLM workflow of the type configurable in our system when it is using a vision language model and prompted to output a JSON configuration file instead of feedback on mathematical homework.

We hope to study these ideas in future work.

\paragraph{Acknowledgments} The authors would like to thank Peter Johnson and the entire \lambdafeedback team for their invaluable support during the implementation of the tool. They also thank Diego Artacho de Obeso for his assistance with human marking.

\printbibliography

\appendix

\section{Model Robustness Analysis: Grading Performance Across Different Evaluators}
\label{app:detailed_correlation_results}

This appendix presents our data on how different workflows/feedback/grading combinations perform in terms of grade correlation with human expert graders and in terms of the \emph{length} of the feedback. The analysis follows the methodology used in the main text, but shows the full results on all the workflow/feedback-model combinations. For the five grading models used (\texttt{gpt-5-mini}, \texttt{gpt-5-nano}, \texttt{gpt-4.1-nano}, \texttt{gpt-4o}, and \texttt{gpt-4o-mini}), we report the range between the best and worst performing grader for each workflow/feedback-model combination, along with the identity of which grading model achieved the best and worst correlation with human expert graders. The grading models were chosen to be relatively lightweight in order to make sure that only feedback that is straightforwardly interpretable will get high correlations. In that sense, the \texttt{gpt-4o} and the \texttt{gpt5-mini} models would a priori be viewed as the most capable grading models tested here. Correlations of human grades with grades generated from human feedback by the grading models are not reported here. They would be expected to be fairly high, as the human feedback is expected to directly justify the human grades.

\subsection{Workflow-Level Comparison}

\begin{figure}[ht]
\centering
\includegraphics[width=\textwidth]{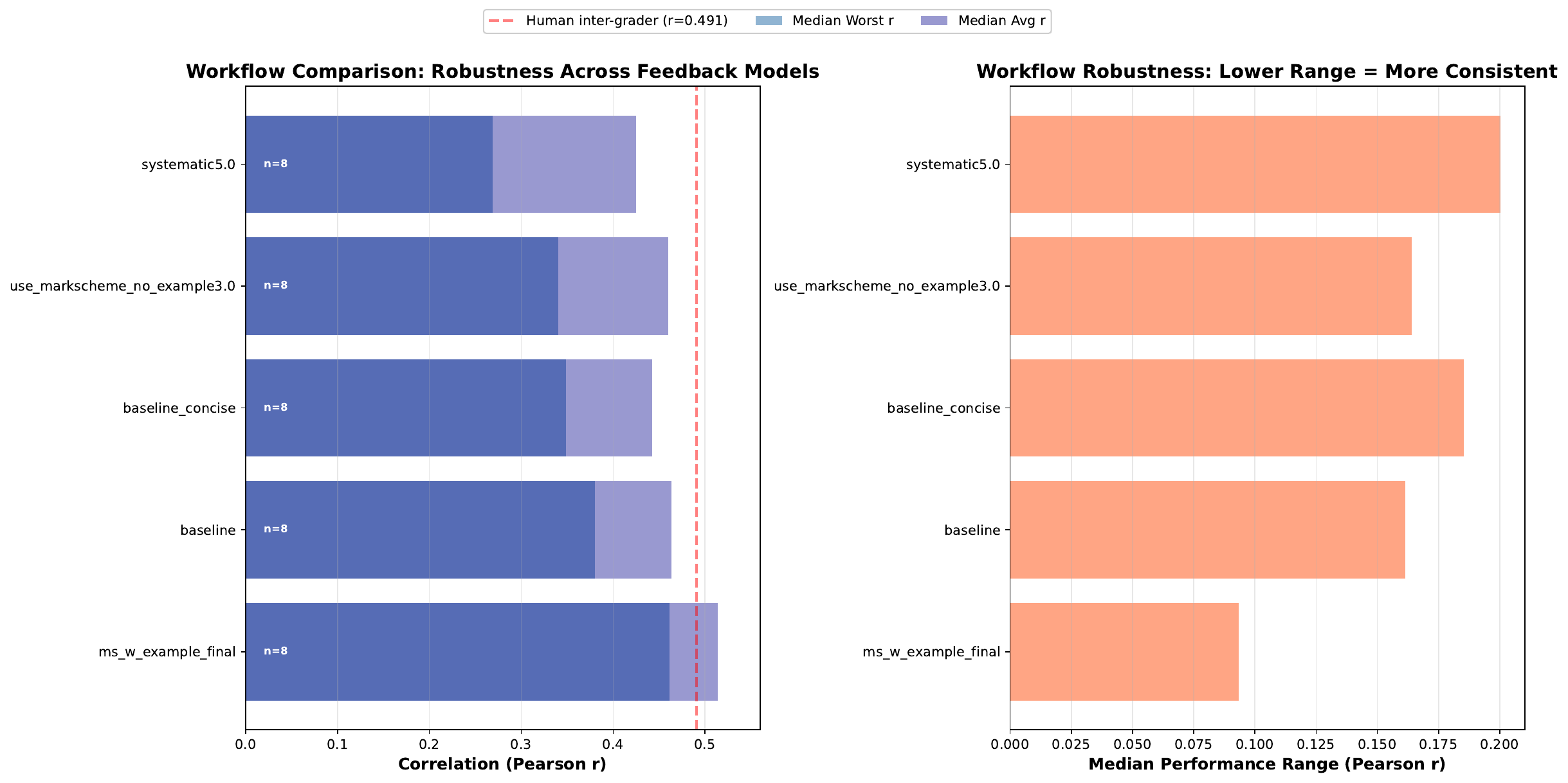}
\caption{Workflow comparison across all feedback models. Left: Median worst-case and average Pearson correlations (higher is better). The red dashed line indicates the human inter-grader correlation baseline (r=0.491). Right: Median performance range (lower indicates more consistent interpretation across graders). The number in each bar (left panel) indicates how many feedback models were tested with that workflow.}
\label{fig:workflow_comparison}
\end{figure}

Figure~\ref{fig:workflow_comparison} compares the five workflows tested, showing both their median worst-case performance (robustness) and median performance range (consistency) independent of the feedback model used. Table~\ref{tab:workflow_summary_complete} provides the detailed statistics.

\begin{table}[ht]
\centering
\caption{Summary of workflow robustness across all feedback models tested.}
\label{tab:workflow_summary_complete}
\footnotesize
\begin{tabular}{lccccccc}
\toprule
\textbf{Workflow} & \textbf{N} & \textbf{Median} & \textbf{Best} & \textbf{Median} & \textbf{Median} & \textbf{Median} \\
 & \textbf{Configs} & \textbf{Worst r} & \textbf{Worst r} & \textbf{Range r} & \textbf{Worst $\tau$} & \textbf{Avg $\tau$} \\
\midrule
ms\_w\_example\_final & 8 & 0.461 & 0.572 & 0.093 & 0.388 & 0.410 \\
baseline & 8 & 0.380 & 0.515 & 0.162 & 0.304 & 0.391 \\
baseline\_concise & 8 & 0.348 & 0.386 & 0.185 & 0.290 & 0.366 \\
use\_markscheme\_no\_example3.0 & 8 & 0.340 & 0.448 & 0.164 & 0.286 & 0.364 \\
systematic5.0 & 8 & 0.269 & 0.422 & 0.200 & 0.219 & 0.347 \\
\bottomrule
\end{tabular}
\end{table}

\textit{Note: N Configs = number of feedback models tested with this workflow; Median Worst r = median worst-case Pearson correlation; Best Worst r = best worst-case correlation achieved; Median Range r = median spread in performance (lower is more robust); Median Worst $\tau$ = median worst-case Kendall's tau.}

The \texttt{ms\_w\_example\_final} workflow demonstrates the strongest overall robustness, with a median worst-case correlation of 0.461 and relatively low median range of 0.093. The \texttt{systematic5.0} workflow shows the weakest robustness with a median worst-case of only 0.269.

\subsection{Feedback Length Comparison}

\begin{figure}[ht]
\centering
\includegraphics[width=\textwidth]{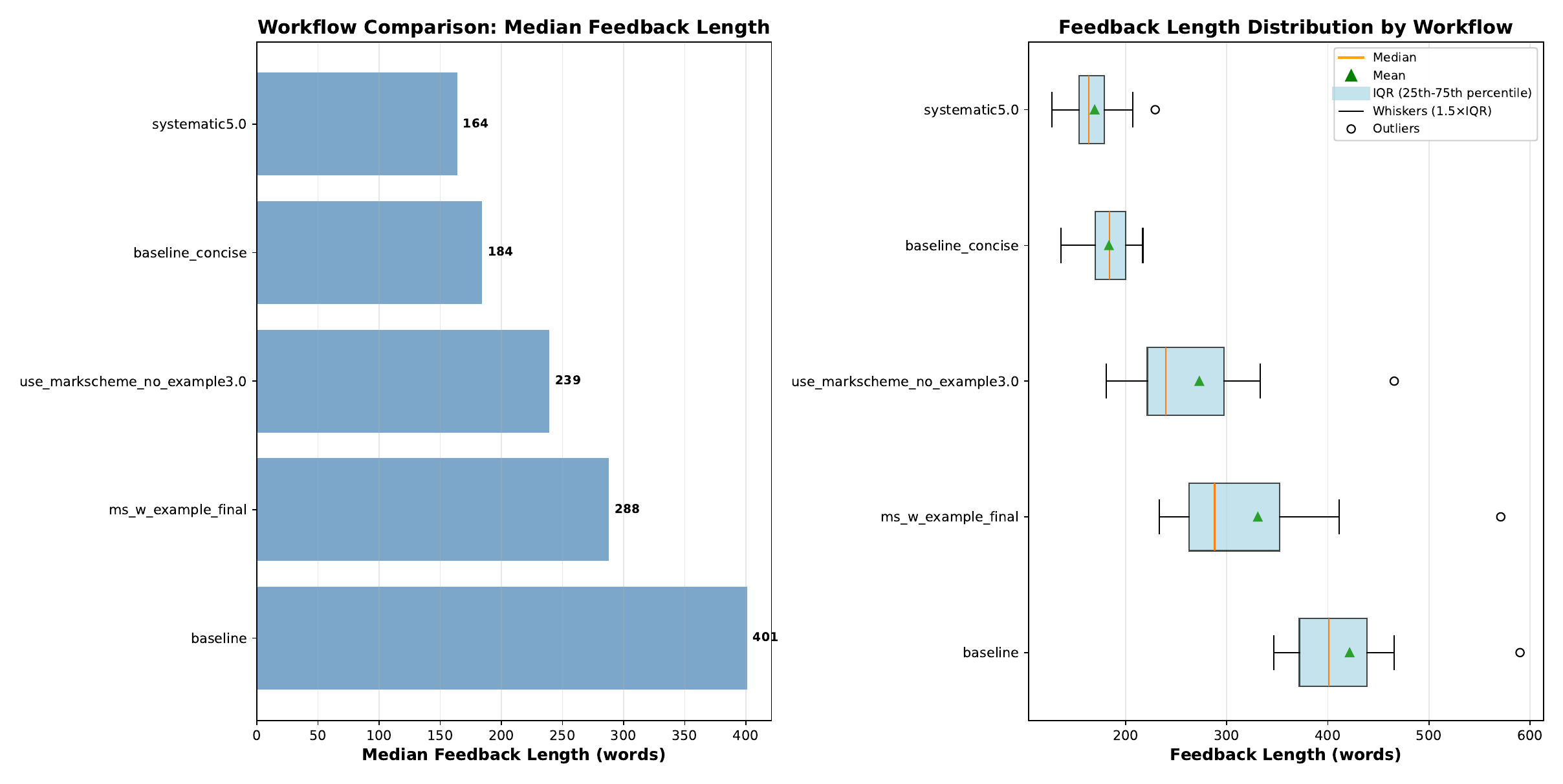}
\caption{Feedback length comparison across workflows. Left: Median feedback length in words for each workflow. Right: Distribution of feedback lengths showing the range and variability. The \texttt{baseline\_concise} workflow produces the shortest feedback (median $\sim$180 words), while \texttt{baseline} and related workflows produce longer feedback (median $\sim$350-400 words).}
\label{fig:feedback_length}
\end{figure}

Figure~\ref{fig:feedback_length} compares the workflows by the average length of feedback generated. Feedback length is an important consideration as it affects both the practical usability of the system (students must read the feedback, and not get overwhelmed by too many details) and for non-reasoning models, the computational cost.

The baseline and baseline-related workflows tend to produce longer feedback (median 350-400 words), while the \texttt{baseline\_concise} workflow produces significantly shorter feedback (median $\sim$180 words). 

The \texttt{ms\_w\_example\_final} workflow, which showed the best robustness, produces moderate-length feedback (median $\sim$260 words), suggesting a balance between detail and conciseness. 

The \texttt{systematic5.0} workflow produces the shortest feedback among the main workflows (median $\sim$160 words), which may contribute to its lower robustness scores as there is less detailed information for grading models to interpret.

\subsection{Performance Ranges for Top Configurations}

\begin{figure}[ht]
\centering
\includegraphics[width=\textwidth]{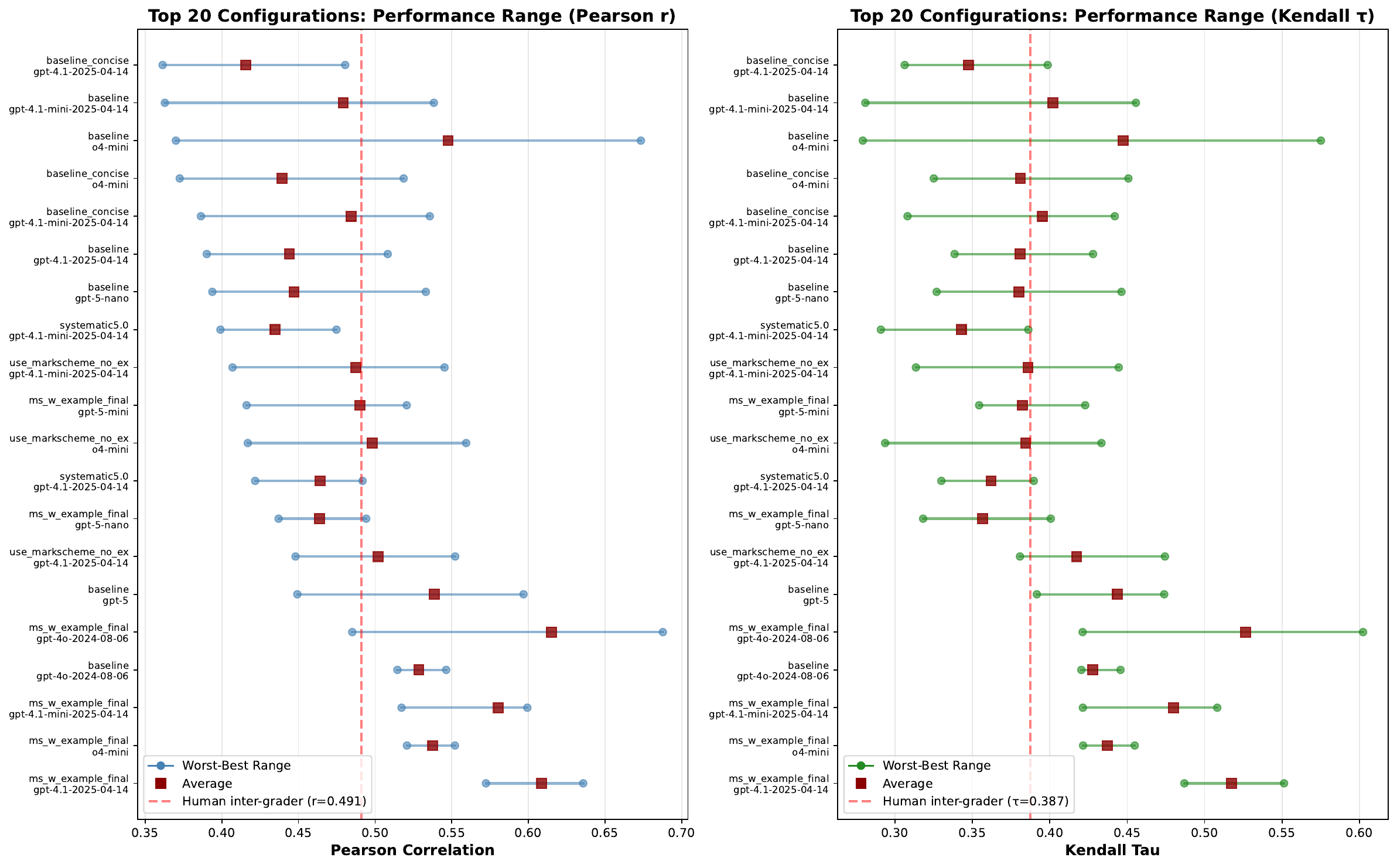}
\caption{Performance ranges for the top 20 workflow/feedback-model combinations. Each horizontal line spans from the worst to best grading model performance, with squares indicating the average. Configurations with shorter lines demonstrate more consistent interpretation across different grading models. Red dashed lines indicate the human inter-grader correlation baseline (r=0.491 for Pearson, $\tau$=0.387 for Kendall).}
\label{fig:performance_ranges}
\end{figure}

Figure~\ref{fig:performance_ranges} visualizes the performance ranges for the top 20 configurations, showing both Pearson correlation and Kendall tau ranges from worst to best grading model.

The visualization shows that the best configuration (\texttt{ms\_w\_example\_final} with \texttt{gpt-4.1-2025-04-14}) maintains a worst-case Pearson correlation of 0.572, with a narrow range of only 0.063, indicating highly consistent interpretability across all grading models.

\subsection{Detailed Results by Feedback Model}

The following tables present detailed performance ranges for each feedback model, showing all workflows tested with that model. For each workflow, we report the worst and best grading model performance (both Pearson r and Kendall $\tau$), the average performance, the identity of the worst-performing grading model, and the range (difference between best and worst Pearson r).

\begin{table}[ht]
\centering
\caption{Performance ranges for feedback model: \texttt{gpt-4.1-2025-04-14}. Each workflow was evaluated by 5 different grading models.}
\label{tab:feedback_gpt_4_1_2025_04_14}
\scriptsize
\begin{tabular}{lcccccccc}
\toprule
\textbf{Workflow} & \multicolumn{3}{c}{\textbf{Pearson Correlation}} & \multicolumn{3}{c}{\textbf{Kendall Tau}} & \textbf{Worst} & \textbf{Range} \\
\cmidrule(lr){2-4} \cmidrule(lr){5-7}
 & \textbf{Worst} & \textbf{Best} & \textbf{Avg} & \textbf{Worst} & \textbf{Best} & \textbf{Avg} & \textbf{Grader} & \textbf{(r)} \\
\midrule
ms\_w\_example\_final & 0.572 & 0.636 & 0.609 & 0.487 & 0.551 & 0.517 & gpt-5-mini & 0.064 \\
use\_markscheme\_no\_example3.0 & 0.448 & 0.552 & 0.502 & 0.381 & 0.474 & 0.417 & gpt-4.1-nano & 0.104 \\
systematic5.0 & 0.422 & 0.492 & 0.464 & 0.330 & 0.390 & 0.362 & gpt-4.1-nano & 0.070 \\
baseline & 0.390 & 0.508 & 0.444 & 0.338 & 0.428 & 0.381 & gpt-4.1-nano & 0.118 \\
baseline\_concise & 0.361 & 0.480 & 0.416 & 0.306 & 0.399 & 0.347 & gpt-4.1-nano & 0.119 \\
\bottomrule
\end{tabular}
\end{table}

\begin{table}[ht]
\centering
\caption{Performance ranges for feedback model: \texttt{o4-mini}. Each workflow was evaluated by 5 different grading models.}
\label{tab:feedback_o4_mini}
\scriptsize
\begin{tabular}{lcccccccc}
\toprule
\textbf{Workflow} & \multicolumn{3}{c}{\textbf{Pearson Correlation}} & \multicolumn{3}{c}{\textbf{Kendall Tau}} & \textbf{Worst} & \textbf{Range} \\
\cmidrule(lr){2-4} \cmidrule(lr){5-7}
 & \textbf{Worst} & \textbf{Best} & \textbf{Avg} & \textbf{Worst} & \textbf{Best} & \textbf{Avg} & \textbf{Grader} & \textbf{(r)} \\
\midrule
ms\_w\_example\_final & 0.521 & 0.552 & 0.538 & 0.421 & 0.455 & 0.437 & gpt-4.1-nano & 0.031 \\
use\_markscheme\_no\_example3.0 & 0.417 & 0.559 & 0.498 & 0.294 & 0.433 & 0.384 & gpt-4.1-nano & 0.142 \\
baseline\_concise & 0.373 & 0.519 & 0.439 & 0.325 & 0.451 & 0.381 & gpt-4o-mini & 0.146 \\
baseline & 0.370 & 0.673 & 0.548 & 0.279 & 0.575 & 0.447 & gpt-4.1-nano & 0.303 \\
systematic5.0 & 0.169 & 0.443 & 0.375 & 0.185 & 0.358 & 0.294 & gpt-4.1-nano & 0.274 \\
\bottomrule
\end{tabular}
\end{table}

\begin{table}[ht]
\centering
\caption{Performance ranges for feedback model: \texttt{gpt-4.1-mini-2025-04-14}. Each workflow was evaluated by 5 different grading models.}
\label{tab:feedback_gpt_4_1_mini_2025_04_14}
\scriptsize
\begin{tabular}{lcccccccc}
\toprule
\textbf{Workflow} & \multicolumn{3}{c}{\textbf{Pearson Correlation}} & \multicolumn{3}{c}{\textbf{Kendall Tau}} & \textbf{Worst} & \textbf{Range} \\
\cmidrule(lr){2-4} \cmidrule(lr){5-7}
 & \textbf{Worst} & \textbf{Best} & \textbf{Avg} & \textbf{Worst} & \textbf{Best} & \textbf{Avg} & \textbf{Grader} & \textbf{(r)} \\
\midrule
ms\_w\_example\_final & 0.517 & 0.599 & 0.580 & 0.421 & 0.508 & 0.480 & gpt-4o-mini & 0.082 \\
use\_markscheme\_no\_example3.0 & 0.407 & 0.545 & 0.488 & 0.314 & 0.444 & 0.386 & gpt-4.1-nano & 0.138 \\
systematic5.0 & 0.399 & 0.475 & 0.435 & 0.291 & 0.386 & 0.343 & gpt-5-mini & 0.076 \\
baseline\_concise & 0.386 & 0.536 & 0.484 & 0.308 & 0.442 & 0.395 & gpt-4.1-nano & 0.150 \\
baseline & 0.363 & 0.538 & 0.479 & 0.281 & 0.456 & 0.402 & gpt-5-mini & 0.175 \\
\bottomrule
\end{tabular}
\end{table}

\begin{table}[ht]
\centering
\caption{Performance ranges for feedback model: \texttt{gpt-4o-2024-08-06}. Each workflow was evaluated by 5 different grading models.}
\label{tab:feedback_gpt_4o_2024_08_06}
\scriptsize
\begin{tabular}{lcccccccc}
\toprule
\textbf{Workflow} & \multicolumn{3}{c}{\textbf{Pearson Correlation}} & \multicolumn{3}{c}{\textbf{Kendall Tau}} & \textbf{Worst} & \textbf{Range} \\
\cmidrule(lr){2-4} \cmidrule(lr){5-7}
 & \textbf{Worst} & \textbf{Best} & \textbf{Avg} & \textbf{Worst} & \textbf{Best} & \textbf{Avg} & \textbf{Grader} & \textbf{(r)} \\
\midrule
baseline & 0.515 & 0.546 & 0.529 & 0.420 & 0.446 & 0.428 & gpt-4.1-nano & 0.031 \\
ms\_w\_example\_final & 0.485 & 0.688 & 0.615 & 0.421 & 0.602 & 0.526 & gpt-4.1-nano & 0.203 \\
baseline\_concise & 0.354 & 0.576 & 0.475 & 0.274 & 0.497 & 0.402 & gpt-4.1-nano & 0.222 \\
use\_markscheme\_no\_example3.0 & 0.349 & 0.523 & 0.470 & 0.311 & 0.447 & 0.401 & gpt-4.1-nano & 0.174 \\
systematic5.0 & 0.331 & 0.524 & 0.447 & 0.311 & 0.434 & 0.385 & gpt-4.1-nano & 0.193 \\
\bottomrule
\end{tabular}
\end{table}

\begin{table}[ht]
\centering
\caption{Performance ranges for feedback model: \texttt{gpt-5}. Each workflow was evaluated by 5 different grading models.}
\label{tab:feedback_gpt_5}
\scriptsize
\begin{tabular}{lcccccccc}
\toprule
\textbf{Workflow} & \multicolumn{3}{c}{\textbf{Pearson Correlation}} & \multicolumn{3}{c}{\textbf{Kendall Tau}} & \textbf{Worst} & \textbf{Range} \\
\cmidrule(lr){2-4} \cmidrule(lr){5-7}
 & \textbf{Worst} & \textbf{Best} & \textbf{Avg} & \textbf{Worst} & \textbf{Best} & \textbf{Avg} & \textbf{Grader} & \textbf{(r)} \\
\midrule
baseline & 0.449 & 0.597 & 0.539 & 0.392 & 0.474 & 0.444 & gpt-4.1-nano & 0.148 \\
ms\_w\_example\_final & 0.296 & 0.524 & 0.439 & 0.206 & 0.404 & 0.329 & gpt-4.1-nano & 0.228 \\
baseline\_concise & 0.237 & 0.496 & 0.397 & 0.203 & 0.448 & 0.330 & gpt-4.1-nano & 0.259 \\
use\_markscheme\_no\_example3.0 & 0.112 & 0.505 & 0.368 & 0.074 & 0.412 & 0.292 & gpt-4.1-nano & 0.393 \\
systematic5.0 & 0.040 & 0.546 & 0.415 & 0.100 & 0.425 & 0.352 & gpt-4.1-nano & 0.506 \\
\bottomrule
\end{tabular}
\end{table}

\begin{table}[ht]
\centering
\caption{Performance ranges for feedback model: \texttt{gpt-5-nano}. Each workflow was evaluated by 5 different grading models.}
\label{tab:feedback_gpt_5_nano}
\scriptsize
\begin{tabular}{lcccccccc}
\toprule
\textbf{Workflow} & \multicolumn{3}{c}{\textbf{Pearson Correlation}} & \multicolumn{3}{c}{\textbf{Kendall Tau}} & \textbf{Worst} & \textbf{Range} \\
\cmidrule(lr){2-4} \cmidrule(lr){5-7}
 & \textbf{Worst} & \textbf{Best} & \textbf{Avg} & \textbf{Worst} & \textbf{Best} & \textbf{Avg} & \textbf{Grader} & \textbf{(r)} \\
\midrule
ms\_w\_example\_final & 0.437 & 0.494 & 0.464 & 0.318 & 0.401 & 0.357 & gpt-5-mini & 0.057 \\
baseline & 0.394 & 0.533 & 0.447 & 0.327 & 0.446 & 0.380 & gpt-4.1-nano & 0.139 \\
use\_markscheme\_no\_example3.0 & 0.320 & 0.476 & 0.415 & 0.278 & 0.357 & 0.328 & gpt-4.1-nano & 0.156 \\
baseline\_concise & 0.157 & 0.490 & 0.362 & 0.134 & 0.417 & 0.306 & gpt-4.1-nano & 0.333 \\
systematic5.0 & -0.010 & 0.473 & 0.328 & 0.020 & 0.395 & 0.253 & gpt-4.1-nano & 0.483 \\
\bottomrule
\end{tabular}
\end{table}

\begin{table}[ht]
\centering
\caption{Performance ranges for feedback model: \texttt{gpt-5-mini}. Each workflow was evaluated by 5 different grading models.}
\label{tab:feedback_gpt_5_mini}
\scriptsize
\begin{tabular}{lcccccccc}
\toprule
\textbf{Workflow} & \multicolumn{3}{c}{\textbf{Pearson Correlation}} & \multicolumn{3}{c}{\textbf{Kendall Tau}} & \textbf{Worst} & \textbf{Range} \\
\cmidrule(lr){2-4} \cmidrule(lr){5-7}
 & \textbf{Worst} & \textbf{Best} & \textbf{Avg} & \textbf{Worst} & \textbf{Best} & \textbf{Avg} & \textbf{Grader} & \textbf{(r)} \\
\midrule
ms\_w\_example\_final & 0.416 & 0.521 & 0.490 & 0.354 & 0.423 & 0.382 & gpt-4.1-nano & 0.105 \\
baseline\_concise & 0.342 & 0.490 & 0.445 & 0.250 & 0.408 & 0.352 & gpt-4.1-nano & 0.148 \\
systematic5.0 & 0.321 & 0.528 & 0.456 & 0.252 & 0.428 & 0.377 & gpt-4.1-nano & 0.207 \\
baseline & 0.260 & 0.515 & 0.434 & 0.239 & 0.412 & 0.350 & gpt-4.1-nano & 0.255 \\
use\_markscheme\_no\_example3.0 & 0.164 & 0.550 & 0.424 & 0.143 & 0.450 & 0.340 & gpt-4.1-nano & 0.386 \\
\bottomrule
\end{tabular}
\end{table}

\begin{table}[ht]
\centering
\caption{Performance ranges for feedback model: \texttt{gpt-4.1-nano}. Each workflow was evaluated by 5 different grading models.}
\label{tab:feedback_gpt_4_1_nano}
\scriptsize
\begin{tabular}{lcccccccc}
\toprule
\textbf{Workflow} & \multicolumn{3}{c}{\textbf{Pearson Correlation}} & \multicolumn{3}{c}{\textbf{Kendall Tau}} & \textbf{Worst} & \textbf{Range} \\
\cmidrule(lr){2-4} \cmidrule(lr){5-7}
 & \textbf{Worst} & \textbf{Best} & \textbf{Avg} & \textbf{Worst} & \textbf{Best} & \textbf{Avg} & \textbf{Grader} & \textbf{(r)} \\
\midrule
use\_markscheme\_no\_example3.0 & 0.331 & 0.503 & 0.450 & 0.271 & 0.390 & 0.343 & gpt-4.1-nano & 0.172 \\
baseline\_concise & 0.325 & 0.641 & 0.472 & 0.306 & 0.556 & 0.420 & gpt-4.1-nano & 0.316 \\
baseline & 0.306 & 0.526 & 0.408 & 0.267 & 0.457 & 0.348 & gpt-4o-mini & 0.220 \\
ms\_w\_example\_final & 0.276 & 0.485 & 0.417 & 0.266 & 0.390 & 0.332 & gpt-4.1-nano & 0.209 \\
systematic5.0 & 0.217 & 0.398 & 0.290 & 0.125 & 0.286 & 0.204 & gpt-5-mini & 0.181 \\
\bottomrule
\end{tabular}
\end{table}

\section*{Appendix: Automated Grading Workflows}

This appendix describes all automated grading workflows evaluated during development. Each workflow processes student mathematical solutions through one or more LLM-based stages to generate feedback and assign a grade from 0 to 5. Directives files are provided in the github repository.

\subsection*{Final Workflows}

The following three workflows were selected for the final performance comparison after testing.

\paragraph{ms\_w\_example\_final}
A three-stage workflow that first creates a question-specific mark scheme by adapting a generic rubric to the particular problem. The LLM then generates feedback by comparing the student's solution against this customized mark scheme. Finally, a third LLM call assigns a grade (0--5) based on the generated feedback and the standard marking criteria.

\paragraph{systematic5.0}
A two-stage analytical approach that first conducts a structured analysis of the student solution, systematically evaluating mathematical rigor, identifying invalid assumptions, and detecting incorrect reasoning. This analysis is then used to generate focused, actionable feedback highlighting specific mistakes and areas for improvement. Grading occurs in a final stage based on the feedback and marking criteria.

\paragraph{use\_markscheme\_no\_example3.0}
Creates a detailed question-specific mark scheme with explicit bullet points describing what constitutes each grade level from 0 to 5. The LLM then evaluates the student solution against this granular rubric to generate actionable feedback. Grading is performed in a separate stage using the generic marking criteria rather than the question-specific scheme.

\subsection*{Baseline Workflows}

\paragraph{baseline}
A minimal single-stage workflow that looks at the question and the student's solution and generates feedback directly. A second step identical to the grading stage of the three final workflows is added to assign a grade based on the generated feedback, the standard marking criteria, the question and the submitted solution.

\paragraph{baseline\_concise}
A minimal single-stage workflow that is identical to the baseline workflow, except that it contains an instruction to limit the feedback to 200 words.

\subsection*{Exploratory Workflows}

These workflows were developed and tested during early iterations but were not included in the final performance comparison.

\paragraph{advanced\_assistant\_test}
Generates question-specific marking criteria from a generic mark scheme template, then analyzes the student solution using these criteria to provide feedback. The workflow emphasizes adapting general principles to specific problem contexts before evaluation.

\paragraph{basic\_config}
A minimal three-stage workflow that extracts the question and solution separately, then generates feedback directly by evaluating the student's work. Grading occurs based on the question and feedback without intermediate analytical steps.

\paragraph{chain\_of\_thought}
Explicitly instructs the LLM to verify each step of the student's solution carefully before writing a single paragraph of final feedback. This approach attempts to improve accuracy through explicit step-by-step verification.

\paragraph{feedback\_with\_example}
Provides the LLM with a concrete example of ideal feedback style (including question, solution, and desired feedback) to guide the tone and structure of feedback generation. This one-shot learning approach aims to standardize feedback quality.

\paragraph{one\_to\_six}
Uses a 1--6 grading scale instead of 0--5 during the evaluation process. The workflow creates a question-specific mark scheme calibrated to this scale, generates feedback, assigns a grade from 1--6, then converts to the standard 0--5 scale through arithmetic transformation.

\paragraph{primary\_solution}
Generates a model solution to the problem, then provides feedback by comparing the student's approach and answer to this reference solution. This allows the LLM to identify deviations from an ideal approach.

\paragraph{simple\_error\_list}
Takes a diagnostic approach by first extracting all errors from the student solution as a structured bullet-point list. Grading is then based directly on this error inventory rather than narrative feedback.

\paragraph{struct\_review}
Produces highly structured feedback organized into four sections: Brief Problem Statement, Weaknesses of Submission, Strengths of Submission, and Advice to Learner. This format provides balanced, actionable guidance in a consistent structure.

\paragraph{systematic}
An early multi-stage analytical workflow that separately identifies erroneous assumptions, logical/conceptual errors, and evaluates solution completeness. These three analyses are synthesized into unified feedback before grading.

\paragraph{use\_markscheme\_no\_example}
Creates a question-specific mark scheme with minimal guidance (defining only 0/5 and 5/5 anchors), allowing the LLM to interpolate intermediate grades. Feedback generation follows the mark scheme, with grading in a final stage.

\paragraph{use\_markscheme\_with\_example}
Similar to \texttt{use\_markscheme\_no\_example3.0} but uses the generic rubric template to guide mark scheme creation. This provides more structure in adapting general principles to the specific question.

\end{document}